\documentclass[useAMS,usenatbib,a4paper, fleqn]{mnras}
\usepackage{ae,aecompl}

\title[Galactic magnetic fields]
{
Magnetic field evolution and reversals in spiral galaxies
}
\author[Dobbs et al.]
{C. L. Dobbs\thanks{E-mail:
dobbs@astro.ex.ac.uk}$^{1}$, D. J. Price$^2$, A. R. Pettitt$^3$, M. R. Bate$^1$ and T. S. Tricco$^1$\\
$^1$School of Physics, University of Exeter, Stocker Road, Exeter, EX4 4QL, UK \\
$^2$Monash Centre for Astrophysics and School of Physics \& Astronomy, Monash University, Clayton, Vic 3800, Australia \\
$^3$Department of Physics, Faculty of Science, Hokkaido University, Sapporo 060-0810, Japan }
\usepackage{amssymb}
\usepackage{amsmath}
\usepackage{upgreek}
\usepackage[pdftex]{graphicx}
\usepackage{epsfig}
\usepackage{multirow} 
\def\aap{{ A\&A}}

\def\apj{{ApJ}}
\def\apjl{{ApJL}}

\def\mnras{{MNRAS}}

\def\nat{Nature}
\def\ssr{{Space Science Reviews}}

\def\apjs{{ApJS}}

\begin{document}
\label{firstpage}
\date{\today}

\pagerange{\pageref{firstpage}--\pageref{lastpage}} \pubyear{2015}

\maketitle

\begin{abstract}
We study the evolution of galactic magnetic fields using 3D smoothed particle magnetohydrodynamics (SPMHD) simulations of galaxies with an imposed spiral potential.  We consider the appearance of reversals of the field, and amplification of the field. We find magnetic field reversals occur when the velocity jump across the spiral shock is above $\approx$20~km~s$^{-1}$, occurring where the velocity change is highest, typically at the inner Lindblad resonance (ILR) in our models. Reversals also occur at corotation, where the direction of the velocity field reverses in the co-rotating frame of a spiral arm. They occur earlier with a stronger amplitude spiral potential, and later or not at all with weaker or no spiral arms. The presence of a reversal at a radii of around  4--6 kpc in our fiducial model is consistent with a reversal identified in the Milky Way, though we caution that alternative Galaxy models could give a similar reversal. We find that relatively high resolution, a few million particles in SPMHD, is required to produce consistent behaviour of the magnetic field. Amplification of the magnetic field occurs in the models, and while some may be genuinely attributable to differential rotation or spiral arms, some may be a numerical artefact. We check our results using {\sc athena}, finding reversals but less amplification of the field, suggesting that some of the amplification of the field with SPMHD is numerical.\end{abstract}


\section{Introduction}
Understanding the magnitude and morphology of galactic magnetic fields is a long-standing problem in galactic astronomy. 
The morphology of galactic magnetic fields is difficult to observe, and poorly understood theoretically. One difficulty is measuring the direction of the magnetic field. Faraday rotation has indicated that the Milky Way contains reversals of the magnetic field (where the magnetic field vector reverses direction), but as yet we do not know whether reversals occur in other galaxies. Another problem is understanding the location of `magnetic spiral arms', where the ordered component of the magnetic field is strongest. In some spiral galaxies, these tend to be aligned with the optical spiral arms as expected (e.g. M51, \citealt{Fletcher2011}) but in other galaxies the ordered component is strongest in the inter-arm regions \citep{Beck2007a}. Many models and simulations use dynamo theory to try and interpret these phenomena, but there is no consensus about the origin of reversals, the differences in morphologies between galaxies, or the physical cause or timescale of magnetic field growth in galaxies.

Most observations indicate there is one reversal of the magnetic field in the Milky Way, in the region of the Sagittarius spiral arm \citep{Frick2001,Nota2010,VanEck2011,Beck2011}. 
\citet{Han2006} suggest there may be numerous reversals within the galactic disc, although most surveys rule out any reversals in the outer Galaxy \citep{Brown2001,VanEck2011}.
Reversals have been observed in a number of simulations, with different types of numerical codes, with and without an explicit turbulent dynamo. Many explanations of reversals involve a turbulent dynamo, and for example, \citet{Chamandy2013} investigate reversals in different types of spiral galaxies starting from a primordial magnetic field strength. Alternatively reversals could be induced by vertical oscillations of the gas which produce a vertical dynamo \citep{Ferriere2000,Gressel2013}. However reversals are also seen in simulations without explicit dynamo terms (e.g. \citealt{Pakmor2013}). Although the reversals seem to readily occur, there is little explanation of what causes the reversal or what properties the reversals depend on.

A further question is how the magnetic field has been amplified to the values seen in low-redshift galaxies, from primordial field strengths. Again, dynamo theory is often invoked to explain amplification of the magnetic field (\citealt{Parker1971}, see also recent review by \citealt{Brandenburg2015}). However, a number of processes may contribute to the amplification of the field including turbulence (as exemplified by standard turbulent dynamo theory), stellar feedback and differential rotation \citep{Gaensler2005}. Recent simulations of isolated galaxies indicate that supernovae feedback can drive a small scale dynamo \citep{Becka2012,Rieder2016}. Differential rotation also amplifies magnetic fields (see simulations by \citealt{Kotarba2009}) although there are doubts that differential rotation alone can be responsible for the present day magnetic field strengths \citep{Brandenburg2015}. In addition to explaining current field strengths, turbulent dynamos have also been explicitly included as an extra term in MHD calculations to reproduce ordered fields (or magnetic spiral arms) between the optical spiral arms \citep{Chamandy2013,Moss2015}.

There are numerous challenges for studying galactic magnetic fields numerically, including the difficulty of modelling magnetic fields in the smoothed particle hydrodynamics (SPH) method, as well as achieving high enough resolution to effectively model the interstellar medium (ISM). 
In our previous work modelling spiral galaxies \citep{DP2008}, we used an SPMHD code whereby the magnetic fields were represented by Euler potentials. The magnetic fields were found to smooth out structure in the disc, and the spiral arms were also seen to induce some measure of disorder in the magnetic field, more prominent in the inter-arm regions. The Euler potentials method had the advantage that the magnetic divergence is zero by construction. However the morphology of the field was limited --- in particular, winding up of the field after multiple rotations of the disc cannot be captured with the Euler potentials. An alternative is to use divergence cleaning methods in SPH to try and limit the value of the magnetic divergence, but these have so far only been applied to galaxy simulations employing relatively low resolutions (e.g. \citealt{Pakmor2013}) or which study only small regions of the ISM. \citet*{Tricco2016} showed that it is possible to model a turbulent dynamo in SPMHD using 3D simulations of a small-scale dynamo in a turbulent, periodic box (representing the ISM). 

Recent grid-based simulations have been used to investigate the role of magnetic fields on gravitational and hydrodynamic instabilities in spiral arms. While magnetic fields were found to have only a small effect, they can limit instabilities seen in purely hydrodynamic models \citep{Lee2014,Kim2015}, and suppress gravitational instabilities in cases with low shear \citep{KOS2002}. A number of grid-based simulations have also self consistently included a galactic dynamo by modelling supernova feedback \citep{Hanasz2009,Kulpa2011}, but again the simulations were relatively low resolution. 

In the present paper, we model isolated spiral galaxies with an imposed spiral potential. Rather than studying the amplification of a primordial field, we start our calculations with magnetic field strengths closer to present day values. Since we observe reversals in our simulations, we investigate their dependence on the galactic potential in order to understand what is required for magnetic field reversals in galaxies. To test the robustness of our results, we examine the evolution of the magnetic divergence, perform resolution tests, vary the strength of the cleaning prescription for the magnetic field, and compare with a grid-based code.

\section{Details of simulations}

We solve the equations of ideal magnetohydrodynamics, given by
\begin{gather}
\frac{{\rm d}\rho}{{\rm d}t} = - \rho \nabla \cdot {\boldsymbol{v}} , \\
\frac{{\rm d}{\boldsymbol{v}}}{{\rm d}t} = - \frac{1}{\rho} \nabla \left( P + \frac{B^2}{2 \mu_0} \right) + \frac{1}{\mu_0 \rho} \nabla \cdot \left( {\boldsymbol{B}} \otimes {\boldsymbol{B}} \right) , \label{eq:momentum} \\
\frac{{\rm d}}{{\rm d}t} \left( \frac{{\boldsymbol{B}}}{\rho} \right) = \left( \frac{\boldsymbol{B}}{\rho} \cdot \nabla \right) {\boldsymbol{v}} , \label{eq:induction} \\
\nabla \cdot {\boldsymbol{B}} = 0 , \label{eq:divbconstraint} 
\end{gather}
where ${\rm d}/{\rm d}t = \partial / \partial t + {\boldsymbol{v}} \cdot \nabla$ is the material derivative, $\rho$ is the density, ${\boldsymbol{v}}$ is the velocity, ${\boldsymbol{B}}$ is the magnetic field, $P$ is the hydrodynamic pressure and $\mu_{0}$ is the permeability of free space.

\subsection{Numerical code and MHD implementation}

Most of our simulations are performed with the smoothed particle hydrodynamics code {\sc sphNG}. The code is based on an original version by  \citet{Benz1990} but has many modifications, those relevant to this paper include individual time steps \citep{Batephd1995} and magnetic fields \citep{Price2004b, Price2004, Price2005, Price2012b, Tricco2012, Tricco2013}. We use the variable smoothing length implementation of SPH, as described in \citet{PM2007}.

Artificial viscosity is included to capture shocks, using a switch \citep{Morris1997} allowing values of $\alpha$ to vary between 0.1 and 1. Using a fixed value of $\alpha=1$ leads to very small differences in the simulations \citep{DobbsIAU2011}. 

The magnetic field is evolved explicitly using the quantity ${\bf B}/\rho$. Stability of the magnetic tension term in the momentum equation (equation~\ref{eq:momentum}) is achieved by using the \citet*{Borve2001} approach (see \citealt{Price2012b}). To deal with discontinuities in the magnetic field, we apply an artificial resistivity \citep{Price2005}, similar to the artificial viscosity used for shocks. In the simulations presented, we use the \citet{Tricco2013} switch for the artificial resistivity with values of $\alpha_B$ varying from 0.1 to~1. Again using a fixed value of $\alpha_B=1$ leads only to small differences in the results. 

The divergence constraint is enforced using constrained hyperbolic divergence cleaning \citep{Tricco2012}, an SPMHD adaptation and improvement of the method by \citet{Dedner2002}. The magnetic field is coupled to a scalar field, which removes divergence error from the magnetic field by propagating $\nabla \cdot {\bf B}$ as a damped wave. The cleaning wave speed, $c_{\rm h}$, is taken to be the fast MHD wave speed, but we also perform one test where we take four times this speed which should make the cleaning more effective (see also \citealt{Bate2014, Tricco2015}). We refer to the multiple of $c_h$ used as the `over-cleaning parameter' henceforth.

For simulations presented in this paper, if the divergence of the magnetic field becomes too large, the time steps become extremely small and the simulation halts. In this paper, this does not occur within the timeframe of our results, but can occur if we run the simulations for longer. In our SPMHD simulations, we do not use the h--averaging method of \citet{Lewis2015}. We found that this method was not necessary, and the resulting asymmetry of the SPH equations led to some irregular results in the case of differentially rotating magnetic discs.

In addition to using {\sc sphNG}, we also performed calculations with the {\sc Athena} grid-based code \citep{Stone2008}, which are presented in the Appendix. We use these results to check the reliability of the {\sc sphNG} results, including the appearance of reversals and field amplification.

\subsection{Galaxy setup and initial conditions}
 We only model the gas component of the disc, using a potential to represent the dark matter halo and stellar component. 
We use a galactic potential, which consists of a logarithmic component to provide a flat rotation curve \citep{Binney1987} and most models also include a time-dependent 2-armed spiral component \citep{Cox2002}. 
In all our calculations we model a gas disc with an outer radius of 10~kpc. Gas is distributed in pressure equilibrium in the vertical direction (see e.g. \citealt{Wang2010}).
Gas particles are initially distributed randomly in the disc, with velocities allocated to the galactic rotation curve plus a 7~km~s$^{-1}$ 3D velocity dispersion. The mean surface density of the gas in each calculation is 8~M$_{\odot}$ pc$^{-2}$. We perform calculations with 1, 4 and 8~million particles, so the corresponding mass of a gas particle is 2500, 625 and 312.5~M$_{\odot}$ in each calculation. The minimum smoothing length is around 2.5 pc for the simulations with 4 and 8 million particles, and around 6~pc with 1~million particles.

All the simulations in this paper are simple isothermal calculations, without self gravity or stellar feedback. In all the simulations the temperature is 100 K. Using higher temperatures tends to smooth out structures in the gas \citep{Dobbs2006}. We initially set up a purely toroidal magnetic field in the disc, with an initial strength of 0.1 $\upmu$G for most of the simulations.  All the simulations were evolved for a minimum of 230 Myr. 

We performed a number of calculations to examine the impact of the galactic potential on the evolution of the magnetic field, including the appearance of magnetic field reversals. The parameters for these calculations are listed in Table~\ref{tab:simtable1}. For our weak shock model (MHDN4Weak), we halve the amplitude of the spiral potential ($p_0$ in \citealt{Cox2002}), and we also performed one calculation with no spiral component of the potential (MHDN4Nosp). These two models, in addition to our fiducial model (MHDN4) are discussed in Sections~\ref{sec:reversal} and \ref{sec:amp} where we present an overview of results with the different galactic potentials. We also performed a calculation to test the effect of starting with a lower magnetic field strength (MHDN4LowB) and a purely hydrodynamic model to examine the impact of the magnetic field (Section~\ref{sec:hydro}). All these simulations employed 4 million particles, except the hydrodynamic model HDN8 which we used to compare with the MHDN8 calculation listed in Table~\ref{tab:simtable2}.

In addition to the simulations mentioned above we also performed several calculations specifically to test the properties of magnetic field reversals. These are listed as the lower four calculations in Table~\ref{tab:simtable1}. We describe the results of these calculations in Section~\ref{sec:further}, which focuses on field reversals. For two of the calculations we varied the pattern speed of the spiral potential, and another model employed a stronger spiral potential.  We also tested one model with a shallower rotation curve (MHDN4Rc2). In this case we varied the form of the logarithmic potential:
\begin{equation}
\phi=V_0^2 \log(R^2+R_c^2+z^2/z_q^2).
\end{equation}
Our fiducial values are R$_c=1$ kpc, $z_q=0.7$ and  $V_0$, the maximum rotation velocity, is 220 km s$^{-1}$.  For the shallower rotation curve, we chose R$_c=2$ kpc which leads to a less steep rotation curve close to the centre of the disc, and produces weaker spiral arms. 

To characterise the different galactic models further, we compute the locations of the inner and outer Lindblad resonances (ILR and OLR) and corotation in Section~\ref{sec:ILR}. Corotation occurs where $\Omega_{sp}=\Omega$ and the ILR and OLR occur where $\Omega\pm\kappa/2=\Omega_{sp}$, where $\Omega$ is the angular velocity of the gas, $\Omega_{sp}$ is the pattern speed (19 km~s$^{-1}$~kpc$^{-1}$ in model MHDN4) and $\kappa$ is the epicyclic frequency. For all models except MHDN4High$\Omega$, corotation lies outside the computational domain of the simulation.

Table~\ref{tab:simtable2} lists the calculations we performed to test the dependence of our results on resolution, and the dependence on the over-cleaning parameter used in the divergence cleaning method. These include a simulation with a higher over-cleaning factor (MHDN4OC4) and simulations with lower (MHDN1) and higher resolution (MHDN8).

\begin{table}
\begin{tabular}{c|cc|c|c|c|c|c}
 \hline 
Run & Spiral & Pattern & R$_c$ & Initial  \\
& amplitude & speed & & field ($\upmu$G) \\
 \hline
MHDN4 &  1 & 1 & 1 & 0.1 \\ 
MHDN4Weak & 0.5 & 1 & 1 & 0.1 \\
MHDN4Nosp &  0 & 1 & 1 & 0.1\\
MHDN4LowB & 1 & 1 &  1 & 0.01 \\
HDN8 & 1 & 1 & 1 & 0\\
MHDN4Strong & 2 & 1 & 1 & 0.1\\
MHDN4Rc2 & 1 & 1 & 2 & 0.1 \\
MHDN4Low$\Omega$ & 1 & 0.5 & 1 & 0.1 \\
MHDN4High$\Omega$ & 1 & 2 & 1 & 0.1\\
\hline
\end{tabular}
\caption{List of calculations performed to test the impact of different galactic potentials on the magnetic field. HDN8 is a purely hydrodynamic model to test the impact of magnetic fields. The top calculation, MHDN4, is our fiducial model. MHDN4Nosp does not include a spiral perturbation to the potential, whilst MHDN4Weak and MHDN4Strong vary the strength of the spiral potential, and MHDN4Low$\Omega$ and MHDN4High$\Omega$ vary the pattern speed. Model MHDN4Rc2 has a shallower rotation curve.}\label{tab:simtable1}
\end{table}

 \begin{table}
\begin{tabular}{c|c|c}
 \hline 
Run & No. &  Overcleaning  \\
& particles & factor  \\
 \hline
MHDN4OC4 & 4 million & 4 \\
MHDN1 & 1 million & 1 \\
MHDN4 & 4 million & 1 \\ 
MHDN8 & 8 million & 1 \\
\hline
\end{tabular}
\caption{List of calculations performed to test the divergence cleaning method and the resolution. MHDN4OC4 has faster divergence cleaning. These models are discussed in Section~\ref{sec:resolution}.}
\label{tab:simtable2}
\end{table}

\section{Results}
\begin{figure*}
\begin{center}
\includegraphics[width=\textwidth]{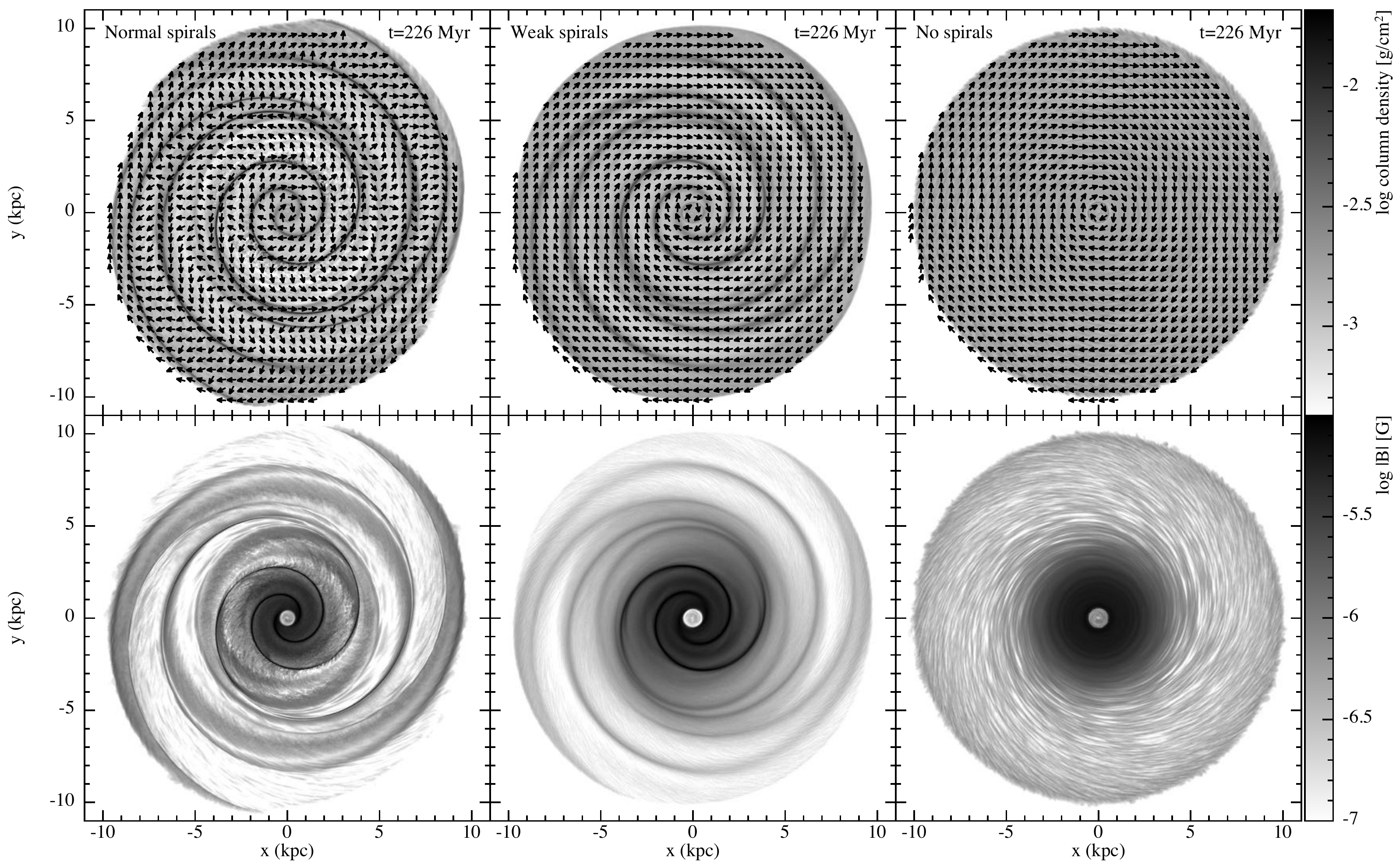}
\caption{Column density of the gas disc (top) together with the orientation (top, arrows) and strength (bottom) of the magnetic field in MHDN4 (left), MHDN4Weak (centre, with a weaker spiral potential) and MHDN4Nosp (right, with no spiral potential). In the fiducial model, there is a reversal of the magnetic field, whereas with no or weak spiral arms there is no reversal at the time shown, 226 Myr. To best show the field morphology and in particular the reversal, the arrows show only the orientation of the field. The lower panels show the magnetic field strength, calculated in a cross section through the midplane. The field grows fastest in the inner 2--3 kpc, saturating at a few ${\mu}G$ after $\sim$150Myr. Magnetic fields are strongest in the inner parts and in the spiral arms.}\label{fig:structure}
\end{center}
\end{figure*}

Figure~\ref{fig:structure} (top panels) shows the structure of the disc, and magnetic field, from our fiducial model (left, MHDN4), a weaker spiral potential (centre, MHDN4Weak) and a model with no spiral arms (right, MHDN4nosp). All calculations use 4 million particles, and are shown at 226 Myr. 
The galactic potential can be seen to have a strong influence on the magnetic field structure. A reversal occurs in the fiducial model (left panel), most evident at a radius of around 4 kpc and extending to radii of 6 or 7 kpc. With a weaker spiral potential (centre panel) the disc produces much weaker spiral arms and no reversals are present at 226 Myr. With no spiral arms (right panel), no reversals are seen in the field. 
The field is also more disordered in the parts of the disc where the field strength is lower, i.e. the inter-arm regions and the outer parts of the fiducial simulation. By contrast, with weak or no spiral arms, the field geometry is regular.
The lower panels of Figure~\ref{fig:structure} highlight the amplification of the magnetic field. All the simulations show amplification of the magnetic field by up to two orders of magnitude in magnetic field strength, particularly in the central regions of the disc and spiral arms.

 \subsection{Magnetic field reversals}\label{sec:reversal}
\begin{figure}
\centerline{\includegraphics[scale=0.38]{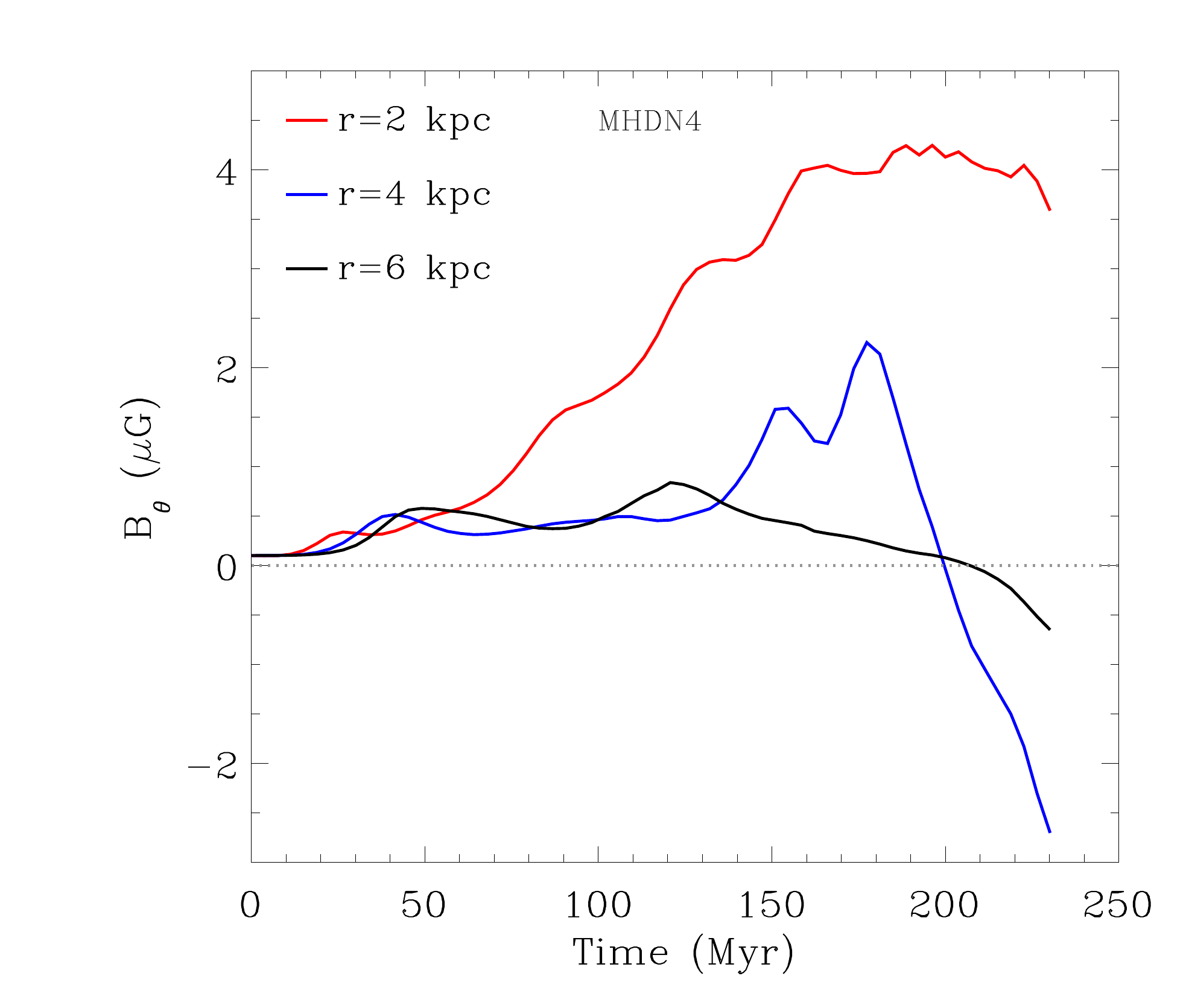}}
\centerline{\includegraphics[scale=0.38]{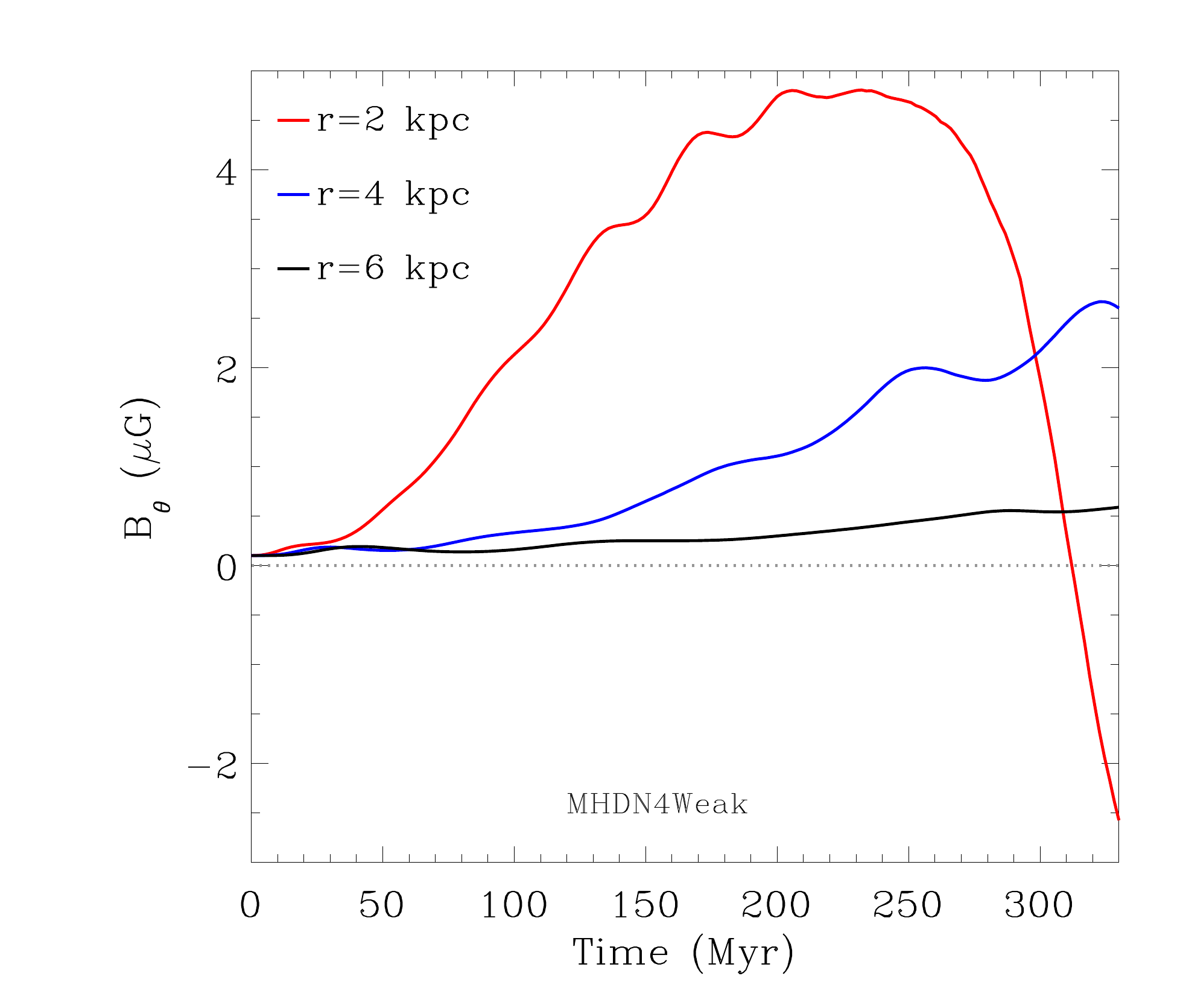}}
\caption{Time evolution of the mean azimuthal magnetic field, comparing our fiducial model (MHDN4, top) to a model with a weaker potential (MHDN4Weak, lower). The change of sign of $\boldsymbol{B}_{\theta}$ in the top panel indicates a reversal at $r \gtrsim 4$kpc after 200 Myr, whilst the model with the weaker potential does not exhibit a reversal (in this case at 2kpc) until after 300 Myr.}\label{fig:btheta}
\end{figure}

\begin{figure*}
\centerline{\includegraphics[width=0.7\textwidth]{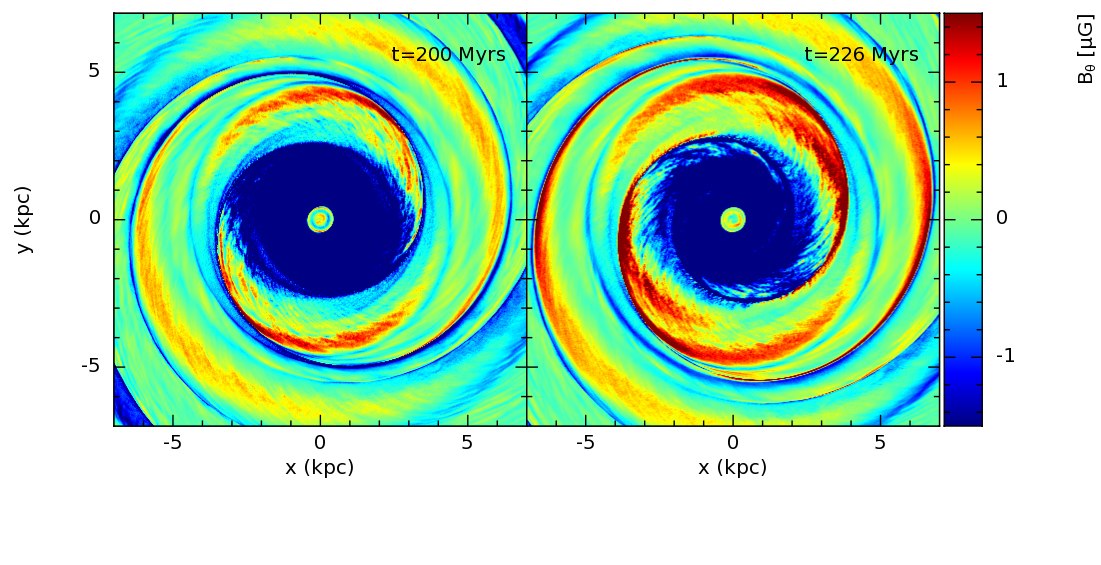}}
\caption{Cross section of $\boldsymbol{B}_{\theta}$ in the midplane of the disc, comparing an earlier (left) to a later (right) time during the reversal which occurs for our fiducial model, MHDN4. The reversal becomes weaker above and below the midplane.}\label{fig:bthetarender}
\end{figure*}

We consider the evolution of $\boldsymbol{B}_{\theta}$ as a proxy for the direction of the field. A change of sign in $\boldsymbol{B}_{\theta}$ indicates a reversal. Figure~\ref{fig:btheta} compares the evolution of $\boldsymbol{B}_{\theta}$ as a function of time in our fiducial calculation (MHDN4, upper panel) to the calculation with a weaker potential (MHDN4weak, lower panel). The figure shows the average of $\boldsymbol{B}_{\theta}$ on the particles computed at three different radii. The average is computed over an annulus of width 200 pc located at the specified radius. In the standard calculation, reversals are seen at the larger radii, after around 200 Myr of evolution. Figure~\ref{fig:bthetarender} shows $\boldsymbol{B}_{\theta}$ rendered at an earlier and later stage of the reversal. At the earlier stage the reversal is predominantly in the inter-arm regions. At the later time, the reversal is also apparent in the spiral arms, and extends from galactic radii of around 3 kpc to 6--7 kpc. With the weaker potential, there is no reversal over the same time frame as the fiducial model, but there is a reversal in the inner part of the disc after 300 Myr. 

\begin{figure}
\centerline{\includegraphics[scale=0.36]{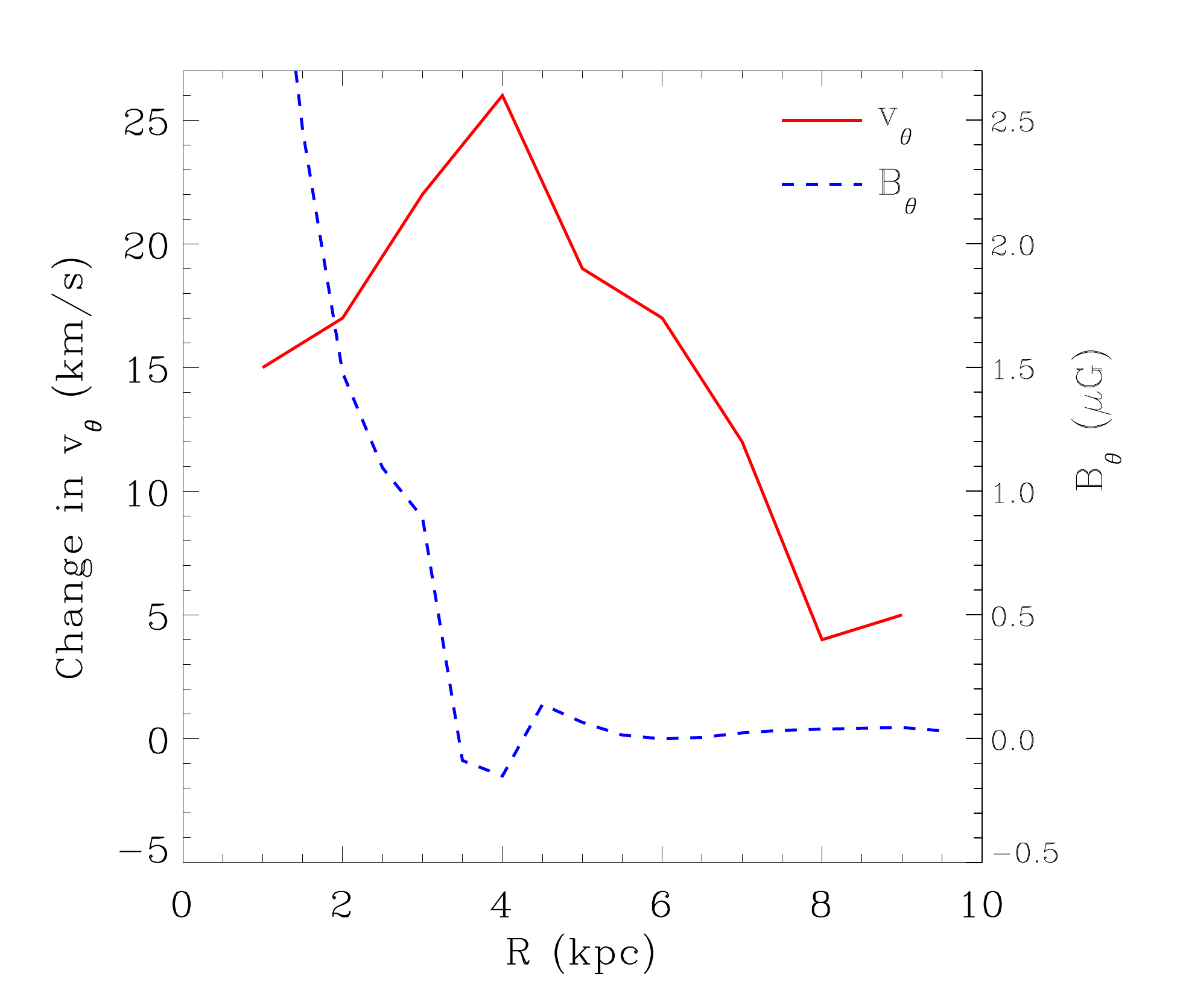}}
\caption{Change in $\boldsymbol{v}_{\theta}$ (solid red line) versus radius for the fiducial model, MHDN4. The greatest velocity change occurs at 4 kpc. The value of $\boldsymbol{B}_{\theta}$ is also shown (blue dotted line) at a time of 211 Myr, just when the reversal has first developed. The reversal (at $R=4$ kpc) is located very close to the maximum velocity change.}\label{fig:vtheta}
\end{figure}

Our simulations thus indicate that the spiral shock has a strong impact on the both the frequency and presence of reversals in the disc. Without spiral arms, reversals do not occur, at least over the timescale of 300 Myr that we consider. In our fiducial model we see a reversal occurring on a timescale of about 200 Myr. With weak spiral arms, a reversal starts to occur about 100 Myr later. Given also that the evolution of $\boldsymbol{B}$ in the limit of ideal MHD will depend on changes in velocity in the gas in the disc, this suggests that the perturbations to the velocity induced by the spiral shock are responsible for the reversals.   

\begin{figure*}
\begin{center}
\centerline{\includegraphics[width=\textwidth]{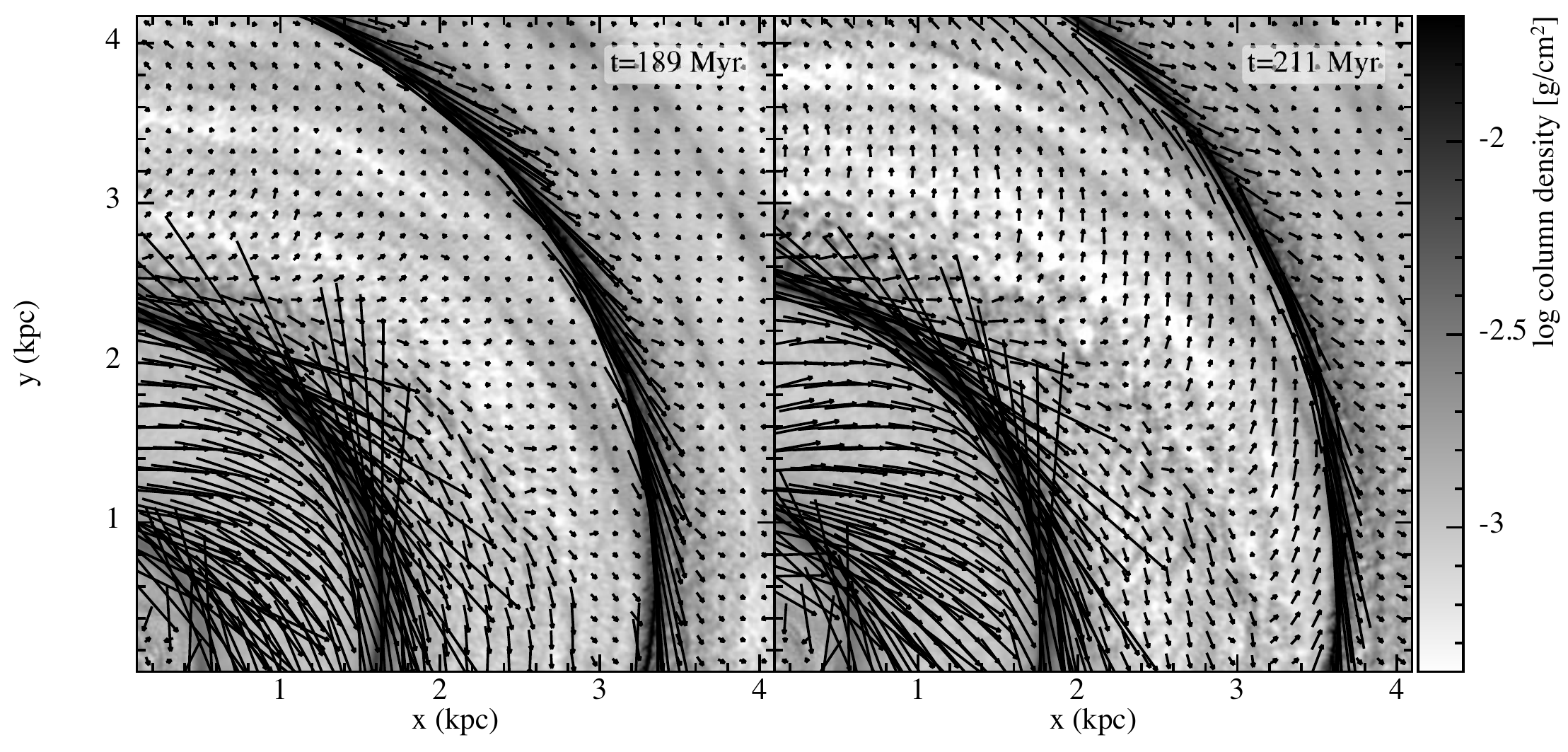}}
\caption{Column density of a section of the disc, shown at times of 189 Myr (left), when a reversal is just starting to occur in the inter arm regions (for $r \gtrsim 2.5$ kpc), and at 211 Myr (right) when the reversal is more clearly established and apparent in the spiral arms. Magnetic field vectors are overplotted on the figures, with the length of each vector indicating the strength of the field.}\label{fig:earlyr}
\end{center}
\end{figure*}

\begin{figure}
\centerline{\includegraphics[scale=0.4]{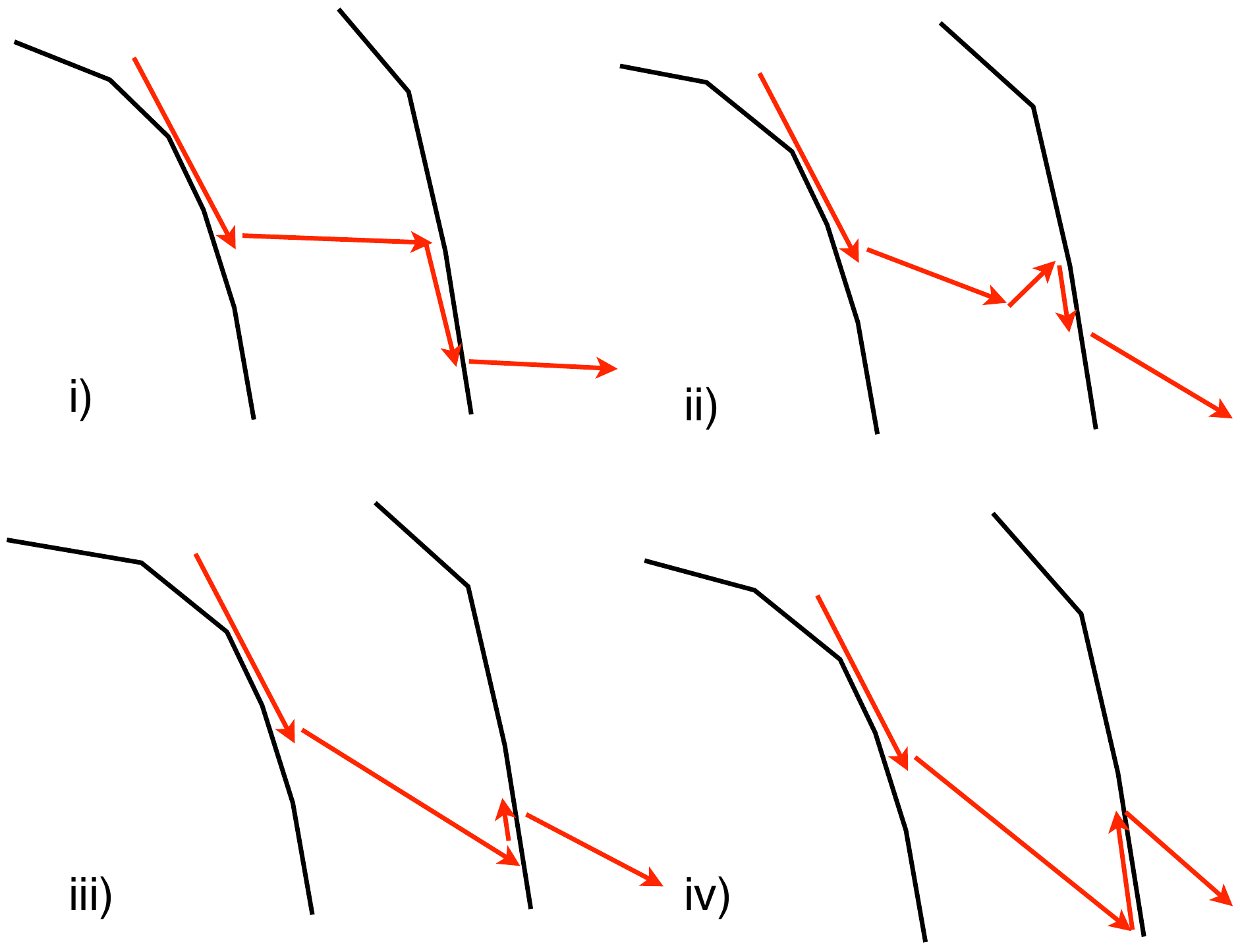}}
\caption{Cartoon of the proposed magnetic field evolution during a reversal. The black lines represent spiral arms and the red arrows the magnetic field. At i) the magnetic field is simply following the velocity field; ii) the magnetic field is trying to straighten out, but still has a kink; iii) the kink is retained as a small reversal in the spiral arm; iv) the magnetic field is amplified in the spiral arm. The size of the arrows does not correspond to the amplitude of the field, rather they simply indicate the magnetic field lines.}\label{fig:cartoon}
\end{figure}

If this is true, then we would expect a reversal to start where the velocity difference is highest. To test this, we calculated the change in $\boldsymbol{v}_{\theta}$ for gas particles at radii of 1--9 kpc (in 1 kpc intervals).  We compute this by first determining the average $\boldsymbol{v}_{\theta}$ in 80 azimuthal bins in a 100 pc width ring situated at each radius. We choose a time of 200 Myr, which is around when the reversal starts to occur. We then determine the difference between the maximum and minimum value of $\boldsymbol{v}_{\theta}$ corresponding to a passage through each spiral arm, and take an average of these two values. Figure~\ref{fig:vtheta} shows the results for our fiducial model MHDN4.
 The largest velocity difference occurs at around 4 kpc. This is indeed where the reversal occurs, as indicated by $\boldsymbol{B}_{\theta}$ in the same figure which becomes negative at around 4 kpc. Figure~\ref{fig:btheta} also indicates that this is where the reversal starts. Figure~\ref{fig:bthetarender} shows that the reversal at 226 Myr occurs between radii of 3 kpc and around 6--7 kpc, which also matches the location where the change in velocity due to the spiral shock is highest. In the model with the weaker potential, the maximum velocity difference at 200 Myr is 13~km~s$^{-1}$, compared to 25~km~s$^{-1}$ in our fiducial model, and no reversal is present in the disc. 

We also examined the morphology of the magnetic field in the vicinity of the reversal, when the reversal first occurs. Figure~\ref{fig:earlyr} shows the magnetic field at 189 Myr (left), when the reversal is just starting. It is evident that the reversal actually starts in the inter arm rather than the arm region. This further confirms our hypothesis that the velocity perturbations to the gas caused by the spiral arms twist the field and lead to reversals. In the spiral arms, the gas flows inwards along the spiral arms (since we are inside corotation). In the inter arm regions, the gas flows in the opposite direction, radially outwards. Figure~\ref{fig:cartoon} shows a cartoon of this idea. If the magnetic field is largely following the velocity field, then there will be significant changes in direction as gas leaves or enters the arm. Figure~\ref{fig:cartoon}i) shows the magnetic field in this case, representing the morphology of the field at earlier times in MHDN4. To reach a more stable configuration, the field may tend to straighten out. With a stronger field in the spiral arm, the field becomes unable to straighten out in the interarm region, and a small reversal occurs. This is shown in Figure~\ref{fig:cartoon} ii), and is similar to the field morphology in Figure~\ref{fig:earlyr}  (left). The reversal may then evolve to the arm region (Figure~\ref{fig:cartoon} iii). Amplification of the field then leads to a stronger reversal in the arm. This is illustrated in Figure~\ref{fig:cartoon}~iv) and corresponds to the morphology in Figure \ref{fig:earlyr} (right). The innermost reversal in the arm occurs at a lower radius compared to the innermost reversal in the inter-arm region in both the cartoon and Figure~\ref{fig:earlyr}. At the larger radii in Figure~\ref{fig:structure}, the field is weaker and more random, and the reversal is less prominent.

In the Appendix, we show some results using {\sc Athena}. We also see reversals using {\sc Athena}, in similar locations to those found here with {\sc sphNG}.

An alternative way to think about the reversals of the magnetic field is in terms of epicycles. In the rest frame of the spiral arms, the gas moves in epicycles, or circular motions. Thus if the magnetic field follows the motion of the gas, then it behaves as a dynamo. It is known that circular motions, and therefore dynamo behaviour can explain galactic magnetic field reversals (e.g. \citealt{Beck2001}) as well as reversals of the solar and terrestrial magnetic fields (e.g. \citealt{Glatzmaier1999,Browning2011}). 

\subsubsection{Further tests of magnetic field reversals}\label{sec:further}
If our interpretation of the field reversals in terms of the velocity perturbations is correct, then the location and frequency of the reversals should be sensitive to the spiral potential. We therefore performed further tests varying the galactic potential. In particular we changed the pattern speed in order to change the corotation radius in the disc as well as the radial location of the maximum velocity difference (MHDN4High$\Omega$ and MHDN4Low$\Omega$). Table~\ref{tab:revtable} lists the location of the reversals in the different models, along with the time of the first reversal and the radius of maximum velocity difference. For the two models MHDN4 and MHDN4Low$\Omega$, the reversal agrees with the location of the maximum velocity change, confirming our hypothesis. For models MHDN4Weak and MHDN4Rc2, the velocity change is smaller and no reversal occurs within 300 Myr. 

The models where we halve and double the pattern speed (MHDN4Low$\Omega$ and MHDN4High$\Omega$) will shift the corotation and Lindblad resonances, and likely also the location of the maximum velocity change, outwards and inwards respectively. We observe that for the MHDN4Low$\Omega$ model, the location of the reversal and the maximum velocity change move outwards, and agree. 
For the model MHDN4High$\Omega$ the origin of the reversal is likely different. The reversal in this case is not located at the radius of maximum velocity difference (there may be an indication of a reversal starting at the edge of the disc but as it is at the outer edge of the disc, it is difficult to be conclusive). However in MHDN4High$\Omega$, the corotation radius is now within the region simulated, and this is the likely cause of this reversal. Since the velocity field changes direction at the corotation radius, with gas moving inwards along the spiral arms inside corotation and outwards outside corotation, this leads to a reversal. The velocity changes within a radius of 7 kpc are small, so there is no reversal further inside the disc.

Our models with only small velocity changes induced by the spiral perturbation (MHDN4Weak and MHDN4Rc2) show no reversals within 300 Myr, although the model MHDN4Weak starts to show a weak reversal after this time. Comparing the velocity differences for models with and without reversals suggests that the spiral potential needs to induce velocity perturbations of around 20 km s$^{-1}$ to induce reversals in the magnetic field.
\begin{table*}
\begin{tabular}{c|c|c|c|c|c|c|c}
 \hline 
Run & Location of  & Location of & Max.  &Time of & ILR & CR & OLR  \\
 & reversal (kpc) & max. $|\nabla v_{\theta}|$ (kpc) & $|\nabla v_{\theta}|$ (km/s) & reversal (Myr) & (kpc) & (kpc) & (kpc) \\
 \hline
MHDN4 & 3.5--4.5 & 4 & 26 & 200 & 4 & 11.6 & - \\
MHDN4Low$\Omega$ & 5--6 & 6 & 23 & 270 & 6 & 23 & - \\
MHDN4High$\Omega$ & 7--7.5 & 9 (edge of disc) & 20 & 200 & None & 5.8 & 9 \\
MHDN4Strong & 3-4 & 3--4 & 27 &120 & 4 & 11.6 & - \\
MHDN4Weak & None & 4 & 13.5 & - & 4 & 11.6 & - \\
MHDN4Rc2 & None & 4 & 16 & - & None & 11.6 & - \\
\hline
\end{tabular}
\caption{Table showing the location of reversals in simulations, and how they compare to the location and magnitude of the maximum velocity gradient and the location of resonances in the disc.
For simplicity, the location of the reversal is measured in the spiral arms rather than the inter-arm region and is determined within the first 10 Myr of the reversal. Due to uncertainty in pinpointing the exact location of the reversals and small changes in the magnitudes of the velocities across the spiral shocks with time, distance values are only accurate to 1 kpc, the velocity values to 1 km/s and the times of the reversals to 10 Myr. The OLRs are not listed apart from for model MHDN4High$\Omega$ as they are well outside the computational domain. The location of the maximum velocity difference is computed at the time of the reversal for the top four models, and at 200 Myr for the lower two models. The model MHDN4Weak develops a weak reversal after 320 Myr.}
\label{tab:revtable}
\end{table*}

Our results allow us to compare three different strength potentials (MHDN4Weak, MHDN4 and MHDN4Strong). For the models with higher spiral potential amplitudes, the reversals occur at the same location in the disc, and with similar velocity changes, but with the stronger potential, the reversal occurs after only around 120 Myr. The model with the weaker spiral arms displays different behaviour. The velocity gradient is weaker at a radius of 3--4 kpc, so a reversal does not occur at this location for the duration of the simulation. The highest velocities occur instead at the latest time of the simulation, at around $R=2$ kpc. Overall the time for the onset of reversals, and likely frequency of reversals, depends on the strength of the spiral shock. 

\subsubsection{Comparison with location of ILR, OLR and CR}\label{sec:ILR}
\citet{Linden1998} interpret reversals in simulated galaxies in terms of the location of the ILR, CR and OLR. Table~\ref{tab:revtable} lists the location of these resonances. Models MHDN4 and MHDN4Low$\Omega$ show an excellent agreement between the location of the ILR, the maximum velocity gradient across the shock, and the location of the reversal. Model MHDN4High$\Omega$ has no ILR or high velocity gradients in the inner part of the disc, and no reversal is observed there. However, there is a higher velocity gradient associated with the OLR and, as mentioned previously, a possible indication of a reversal at large radii. The main reversal is in mild conflict with the location of corotation.  However this discrepancy occurs at least partially because the arms are weak at corotation (there is also a discontinuity in the arms at this point). The reversal becomes clearer when the arms become stronger further away from corotation, which is the location noted in Table~\ref{tab:revtable}.

\subsubsection{Comparison with previous literature}
Our findings in this section are consistent with previous work, particularly \citet{Linden1998}, who performed calculations of barred spiral galaxies. They did not use a hydrodynamical scheme, instead employing an N-body code with a cloud collision scheme coupled to the MHD evolution component of the \textsc{zeus} code. They likewise found a reversal at corotation and also at the Inner and Outer Lindblad resonances. The morphology of the field in their simulations was similar to the field geometry in our simulations, in particular the presence of reversals arising in the inter-arm region. Some reversals were seen within the spiral arms (the field changed direction twice within the spiral arms) which are not observed in our calculations, most likely because the spiral arms were wider in \citet{Linden1998}. Similar to the present paper, they interpreted reversals in terms of the change in direction of velocity at corotation and higher velocity gradients in the vicinity of the ILR and OLR. 
Magnetic field reversals have also been noted in more recent MHD simulations, using both \textsc{AREPO} and \textsc{RAMSES} \citep{Pakmor2013,Dubois2010}. 

Calculations have also investigated the influence of galactic potentials with an explicitly included dynamo \citep{Poezd1993,Chamandy2013}, over much longer timescales (Gyrs). These studies found less influence of the spiral arms on the morphology of the field \citep{Chamandy2013}, but noted that the effect of spiral forcing is limited to enhancing the `$\alpha$-effect' in the spiral arms. The rotation curve, which we have demonstrated to have an impact on the magnetic field, was found to be an important factor in the growth rate of the dynamo and subsequently the evolution of reversals in dynamo models \citep{Poezd1993}.

\subsubsection{Comparison with Milky Way}
The Milky Way is the only galaxy where there is evidence of a reversal in the magnetic field \citep{Beck2011}, although it is difficult to measure reversals in external galaxies. Here we define a reversal as occurring where the magnetic field lines move outwards, rather than inwards along the spiral arms. The determination of the sign of the field in the Milky Way comes from Faraday rotation measures from polarised emission typically from pulsars, as well as extragalactic sources \citep{Vallee2005mhd,Han2006,Han2007}. Although the locations of reversals are unclear, and the direction of the field uncertain for much of the Galaxy, according to Figure~7 of \citet{Beck2011} (see also \citealt{Vallee2005mhd}), there is a fairly certain reversal in the nearby Sagitarrius arm. The location of this reversal is similar to the one in our models, which occurs around radii of 4--6 kpc. 

Although our model shows agreement with the Milky Way, the locations of reversals in our models depend on the location of corotation and the ILR and OLR and/or the greatest change in velocity across the spiral arms. The potential we use is based on the Milky Way \citep{Cox2002}, however the dynamical model of the Galaxy, and locations of resonances are not well known. It may even be that our Galaxy is a flocculent spiral without clear corotation or ILR radii \citep{Baba2013,Pettitt2015}. If our Galactic model is correct, then one potential discrepancy is that we might expect a reversal at larger radii due to corotation, which contradicts observations. Alternatively, the Galaxy may be better represented by a model where corotation lies at the end of the bar (which we do not model here), and corotation is associated with the known reversal. Then there may be either no resonances, or no large velocity gradients in the outer Galaxy. This could potentially fit better with the observational picture presented in \citet{Beck2011}, and agree with recent observed peculiar motions of maser sources in spiral arms. No significant peculiar motions are found in the Sagittarius arm \citep{Wu2014}, whilst other results indicate deviations from Galactic rotation are at most $\sim8$~km~s$^{-1}$ in other spiral arms \citep{Choi2014,Sato2014,Hachisuka2015}.

\begin{figure*}
\includegraphics[width=0.8\textwidth]{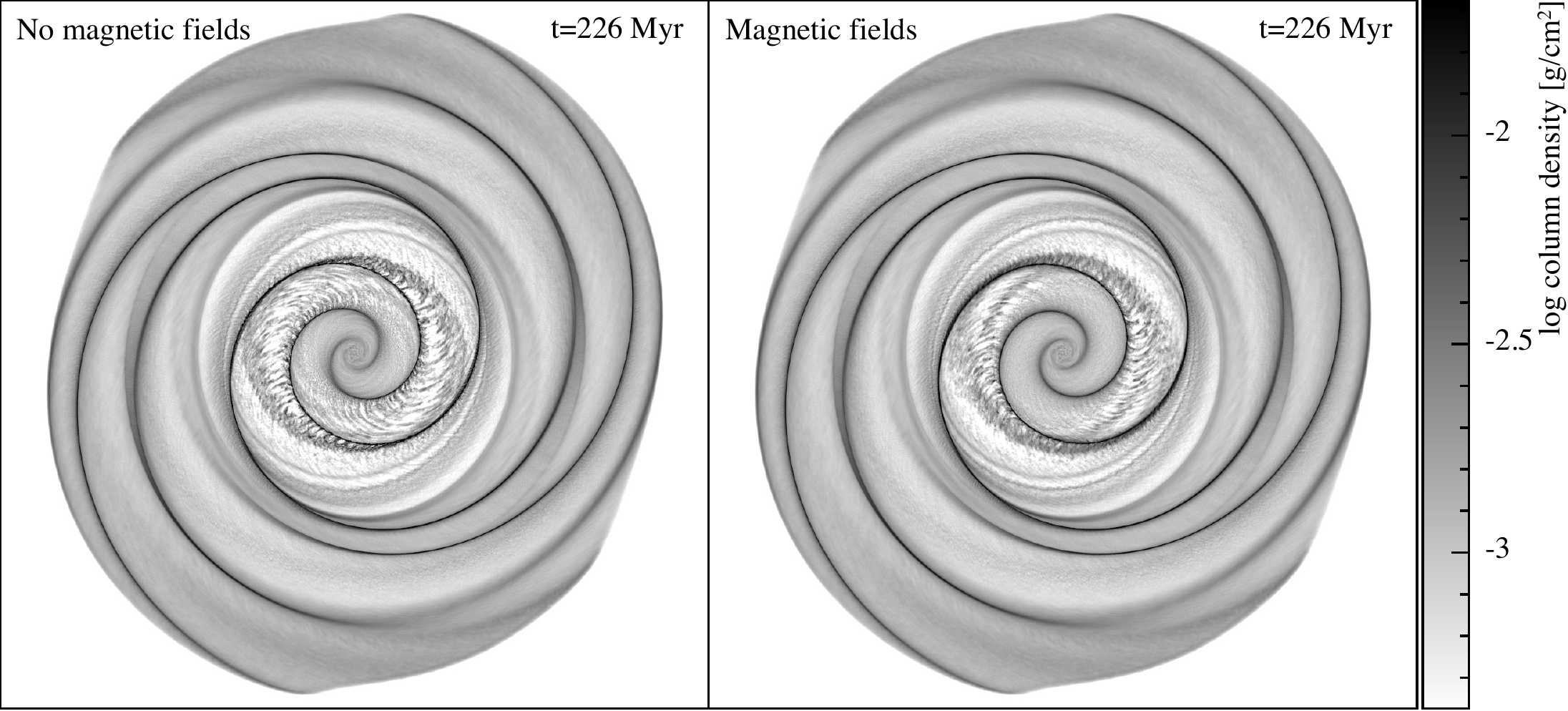}
\caption{Column density maps for simulations without (left) and with (right) magnetic fields. The model on the right is the high resolution magnetic field model (MHDN8). The substructure in the disc is slightly more defined without magnetic fields. }\label{fig:nonmhd}
\end{figure*}

\subsection{Comparison with and without magnetic fields}\label{sec:hydro}
 We repeated our highest resolution calculation, MHDN8 with 8 million particles, but without magnetic fields (HDN8 in Table~\ref{tab:simtable1}). Figure~\ref{fig:nonmhd} shows column density maps of the two calculations at 226 Myr.  The magnetic field has surprisingly little impact on the structure of the disc --- the large scale structure appears similar in both calculations. Some of the substructure is smoother, or reduced, with magnetic fields. The magnetic field induces an additional pressure which smoothes out some of the substructure. This effect of the magnetic pressure smoothing out structure was also found in our previous simulations with Euler potentials \citep{DP2008}. This finding is in qualitative agreement with \citet{Lee2014} and \citet{Kim2015}, who used a grid-based code but also found that substructure was reduced when including magnetic fields.

\subsection{Amplification of the magnetic field}\label{sec:amp}

\begin{figure}
\centerline{\includegraphics[scale=0.38]{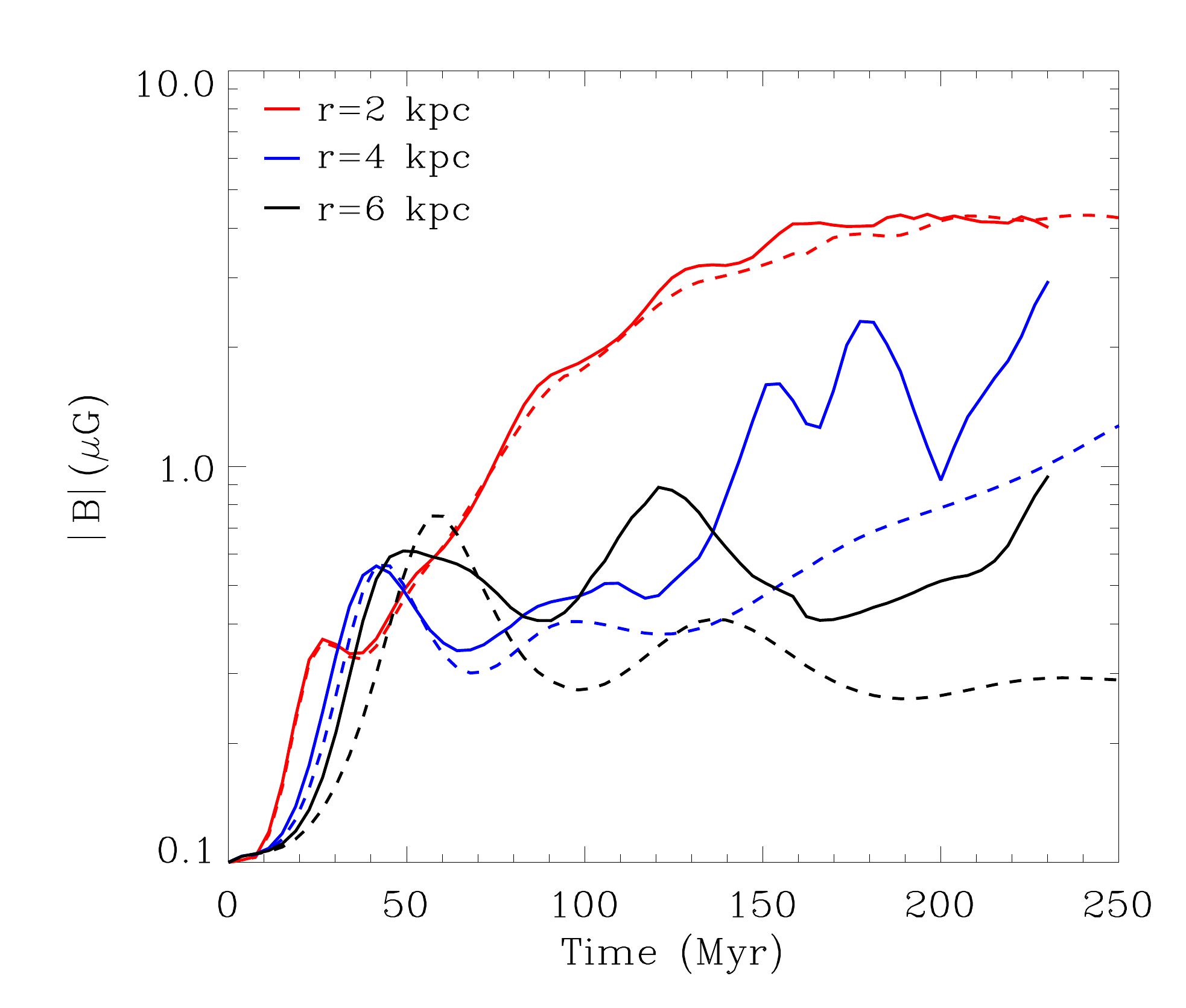}}
\caption{Time evolution of the mean total magnetic field strength at radii of 2, 4 and 6 kpc, comparing simulations with (solid lines, MHDN4) and without spiral arms (dashed lines, MHDN4Nosp).}\label{fig:btot}
\end{figure}
As mentioned in the Introduction, there is some uncertainty about the expected behaviour of magnetic fields in galactic discs. \citet{Kotarba2009} show amplification of the magnetic field by spiral arms. However \citet{Stas2015} suggest that amplification of the field in their SPMHD simulations is due to noise in the initial conditions.  Figure~\ref{fig:btot} shows the time evolution of the magnetic field strength for the calculations with and without the spiral potential (MHDN4 and MHDN4Nosp), computed at three different radii, averaged over an annulus of width 200 pc in each case. The results indicate that the field strength in the disc saturates at $\sim 4\upmu$G. Although the results in Figure~\ref{fig:btot} are only shown up to 250 Myr, for the no-spiral case the field strength saturates at a few $\upmu$G at $r=2$ kpc while slowly increasing at the larger radii until past 300 Myr. Figures~\ref{fig:structure} and \ref{fig:btot} also demonstrate that the strongest amplification of the magnetic field occurs in the central region of the disc. The field strength in the central part of the disc does not depend on the presence of spiral arms. That is, the spiral arms appear to influence the magnetic field only for $r \gtrsim 4$kpc. The field strength in the outer disc is approximately twice as large with spiral arms compared to without.

Amplification of the magnetic field is expected in a differentially-rotating disc given some perturbation in the velocity field. Here we take a similar approach to \citet{Zweibel1987} and simply consider the MHD equation for the time evolution of the magnetic field,
\begin{equation}
\frac{\rm{d}\boldsymbol{B}}{\rm{d}t}=\nabla \times (\boldsymbol{V} \times \boldsymbol{B}), \label{eq:induction2}
\end{equation}
where $\boldsymbol{V}$ is the velocity of the disc. Consider the case where the disc has only circular velocities, with no variation in the $z$-direction (true for our galaxy setup) and thus $\boldsymbol{V}=\boldsymbol{V_c}(r)$ (although in reality there will be some perturbation from circular velocities in the simulations in the azimuthal direction). Then $\boldsymbol{V_c}=r \Omega(r) \boldsymbol{\theta}$ where $\Omega(r)$ is the angular velocity of the disc. Then Equation~\ref{eq:induction2} can be used to give an estimate of the amplification of the radial component of the magnetic field according to
\begin{equation}
\frac{\rm{d}\boldsymbol{B}_{\theta}}{\rm{dt}}=r \boldsymbol{B}_{\theta} \frac{\rm{d} \Omega}{\rm{d}r}.
\end{equation}
Thus we would expect a linear increase in $B_{\theta}$. Figure~\ref{fig:btheta} shows $B_{\theta}$ as a function of time. We observe a slow increase in $B_{\theta}$ for the larger radii, but a superlinear increase at the smallest radius. Thus at the larger radii the field amplification is more consistent with theory, whereas at $r=2$ kpc, the amplification is inconsistent. This picture fits with the finding in Figure~\ref{fig:btot} that at $r=2$ kpc, the amplification is independent of the spiral arms. Probably at these small radii, the amplification may be driven by perturbations of the initial conditions from exact equilibrium. Even when in vertical equilibrium, it is difficult to entirely eliminate structure (typically low amplitude rings, see also \citealt{Few2016}) in the centre of the disc (whilst at larger radii there is no such structure). Thus in the centre of the disc at least, field amplification may be driven mostly by small perturbations at the resolution scale (as supposed by \citealt{Stas2015}). 
Figure~\ref{fig:btot} shows that there is some amplification due to the spiral structure at larger radii (as supposed by \citealt{Kotarba2009}), but it is not particularly substantial, only a factor of 2 or so. In the grid code simulations, we see only a small amplification of the field, again suggesting that the large amplification seen at least at $r=2$ kpc is at least partly numerical.

As well at the total magnetic field, and $B_{\theta}$, we also looked at the evolution of $B_r$ (e.g. Figure~\ref{fig:cleaning}). We would expect, given a purely toroidal initial magnetic field, $B_r$ to remain very small. We find $B_r$ grows to a small ($\sim$10\%) but non-negligible fraction of the total field. This growth of $B_r$, in this case across the disc, is likely a numerical artefact, and we would not expect that $B_r$ should be amplified to these levels. Indeed in the grid-based calculations, $B_r$ remains negligible (whereas growth in $B_{\theta}$ is non-negligible). Increasing the efficiency of divergence cleaning does slightly affect the $B_{r}$ in the inner few kpc (lower panel of Figure~\ref{fig:cleaning}), but has only a small effect on the overall field amplification (top panel of Figure~\ref{fig:cleaning}). \citealt{Pakmor2013} found even stronger field amplification, but did not employ any divergence cleaning aside from the `Powell scheme' which merely preserves divergence errors (and as a result their divergence errors are 1--2 orders of magnitude larger than in our calculations).

\section{Numerical tests}\label{sec:resolution}
In this section we present tests of the resolution and divergence cleaning method with {\sc sphNG}. In the Appendix, we present results with {\sc Athena}.
\begin{figure}
\centerline{\includegraphics[scale=0.3]{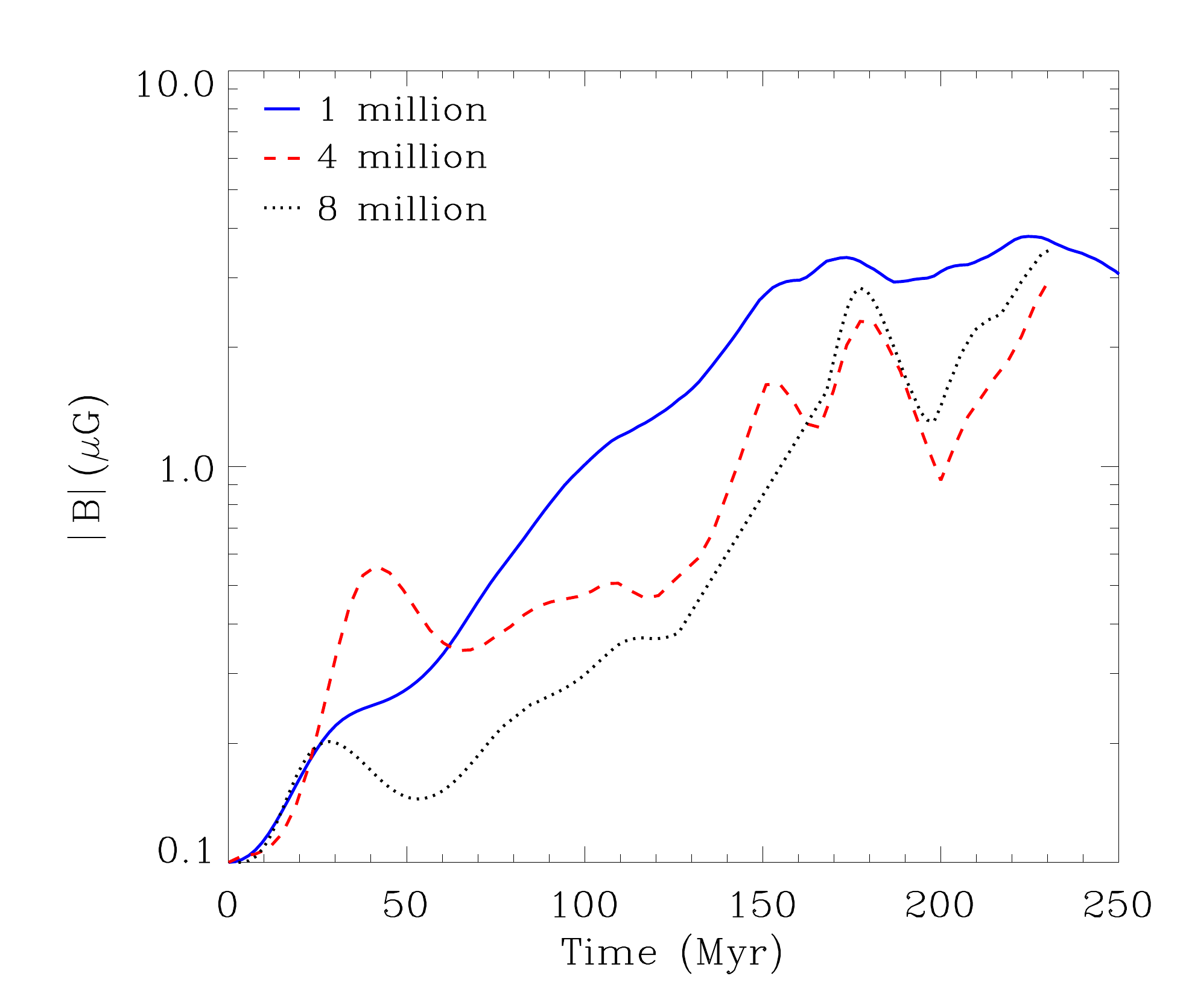}}
\centerline{\includegraphics[scale=0.31]{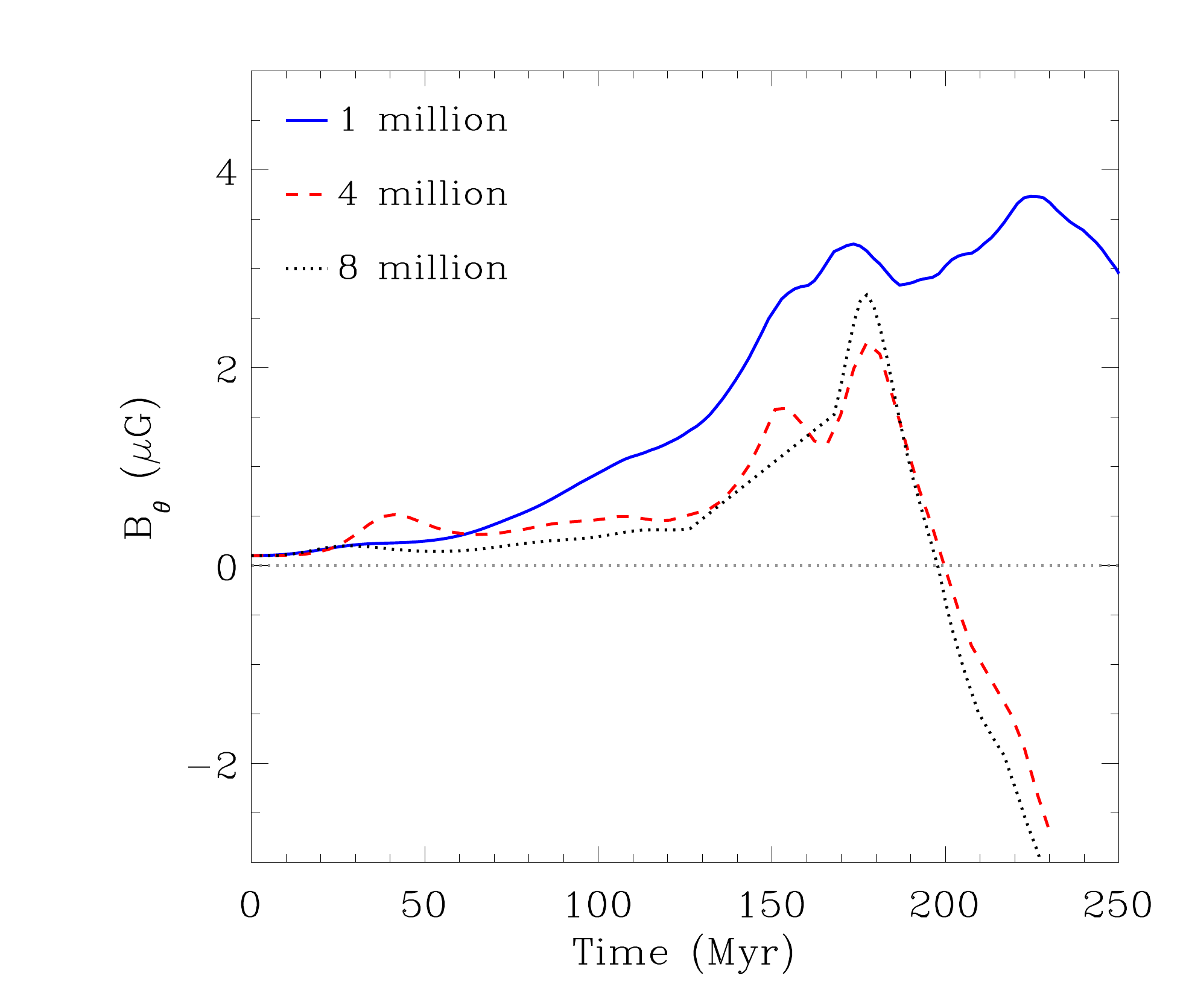}}
\caption{Time evolution of $|\boldsymbol{B}|$ (top) and $\boldsymbol{B}_{\theta}$ (lower) for simulations at different resolution (MHDN1, our fiducial model, MHDN4 and MHDN8). The evolution is shown for a ring located at a radius of 4 kpc in all the above plots. In all cases there is much better agreement between the simulations with 4 and 8 million particles compared to the simulation with 1 million particles.}\label{fig:bresolution}
\end{figure}
\subsection{Resolution tests}
\begin{figure*}
\begin{center}
\includegraphics[width=\textwidth]{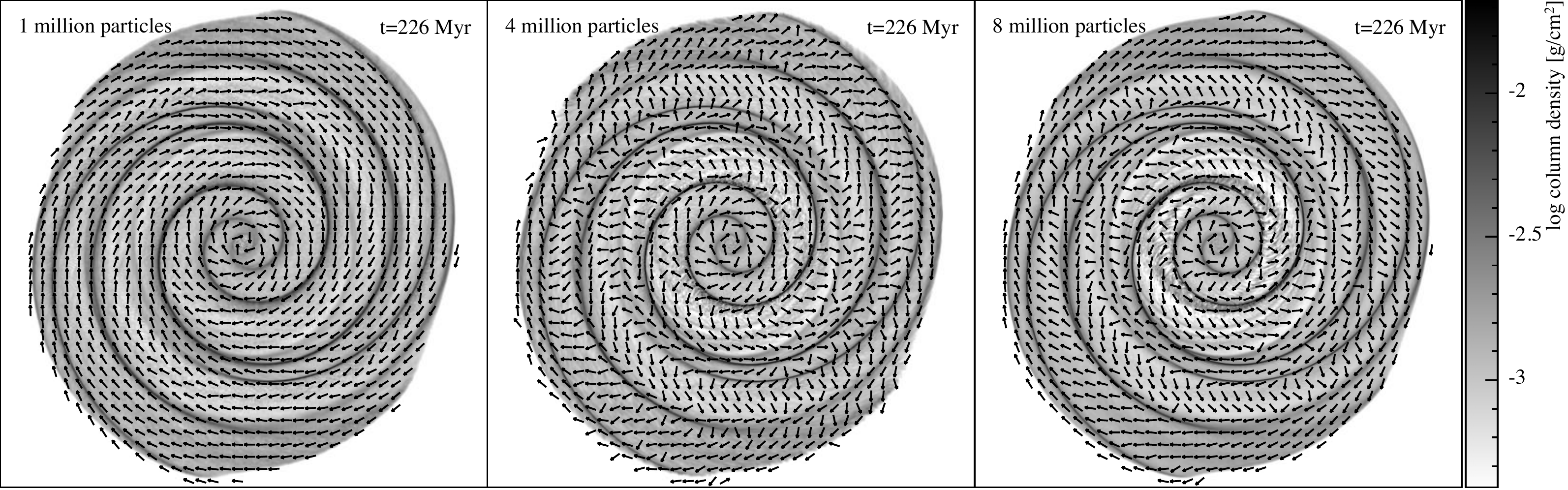}
\includegraphics[width=\textwidth]{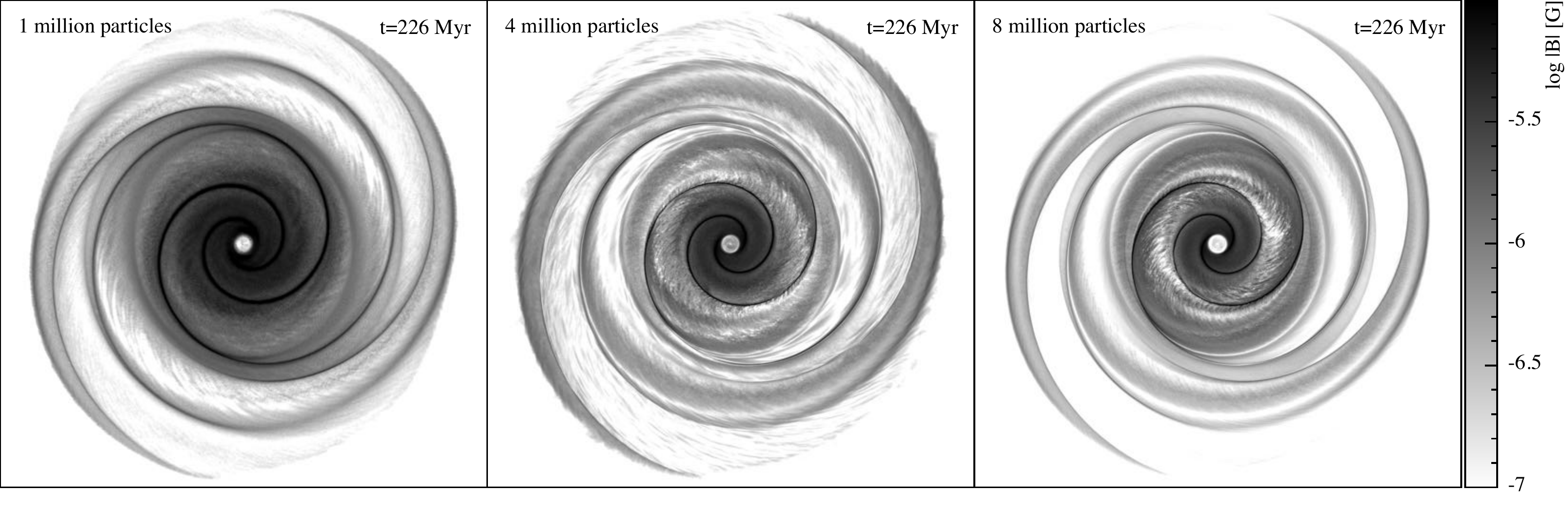}
\caption{Column density maps and magnetic field vectors (top), and midplane magnetic field strength (bottom) in the simulations with 1 (left), 4 (middle) and 8 (right) million particles. The higher resolution simulations show more substructure. The simulations with 4 and 8 million particles produce field reversals although the simulation with only 1 million particles contains no reversal at this stage.}\label{fig:resolution}
\end{center}
\end{figure*}

\begin{figure*}
\begin{center}
\includegraphics[width=0.7\textwidth]{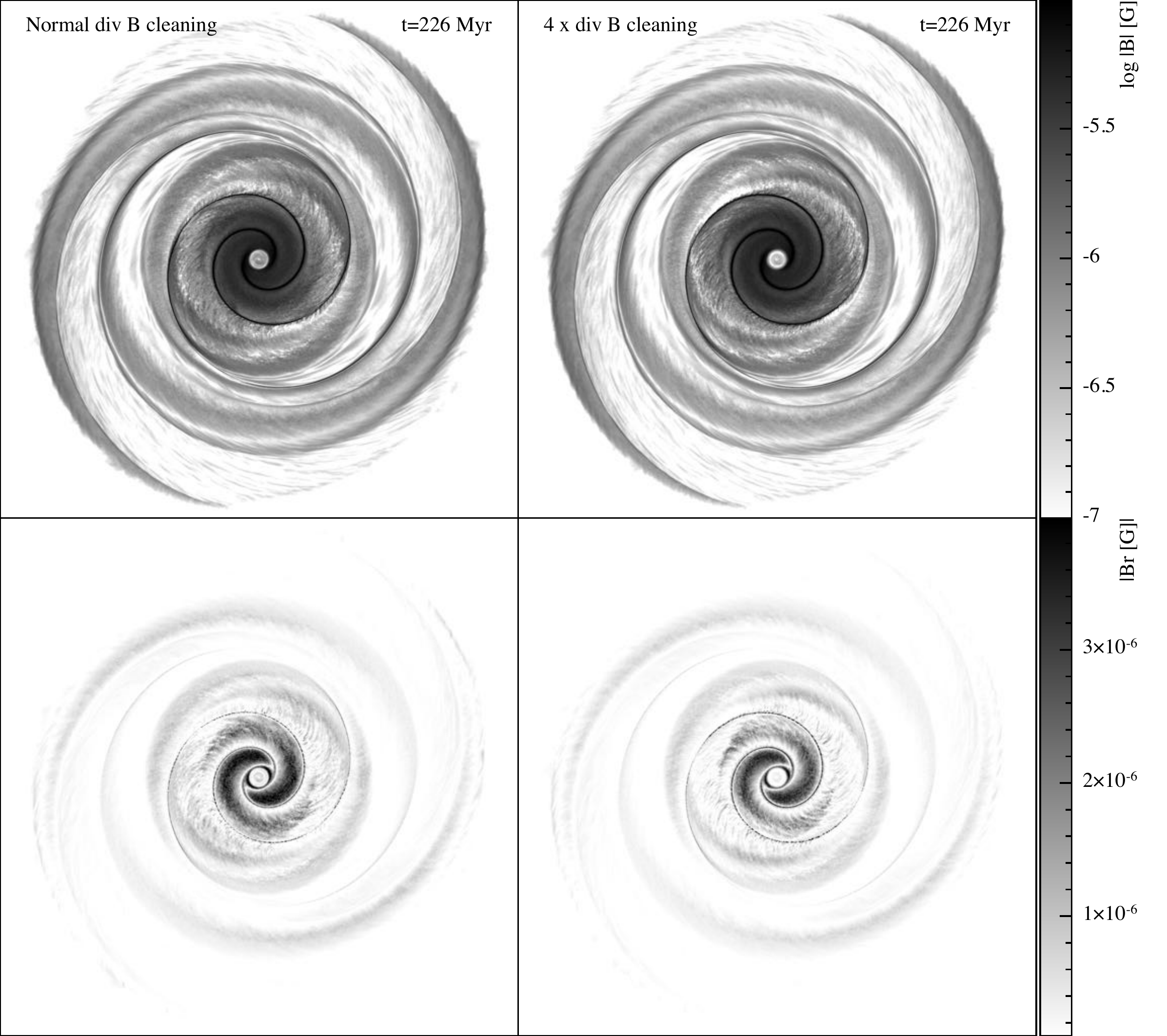}
\caption{(Top) Midplane magnetic field strength in calculation MHDN4 (left) compared to MHDN4OC4 (right, where the over-cleaning is a factor of four higher). (Bottom) Radial component of the field in the midplane, shown with linear scale. The over-cleaning factor only has a small effect on the disordered field in the inner few kpc. The reversals of the magnetic field are not affected by divergence cleaning.}\label{fig:cleaning}
\end{center}
\end{figure*}

We performed resolution tests with 1 and 8 million particles (MHDN1 and MHDN8). Figure~\ref{fig:resolution} shows the structure of the disc and magnetic field at 226Myr for these calculations compared to MHDN4. It is clear from Figure~\ref{fig:resolution} that the substructure in the disc increases with increasing resolution. In particular there is visibly more structure around the ILR at higher resolution, which is not evident at lower resolution. 
This is presumably a combination of higher resolution resolving better the small-scale structure, and extra pressure from the magnetic field at lower resolution, since the magnetic field strength is slightly higher (see Figure~\ref{fig:bresolution}).
With only 1 million particles, the disc is comparatively smooth.

There are also differences in the magnetic field. In the calculation with only 1 million particles, there is no indication of a field reversal (a reversal does occur but not until another 50 Myr). Both the calculations with 4 and 8 million particles show a reversal, although the reversal occurs at slightly smaller radii with 8 million particles, and the field is slightly stronger in the spiral arms. The latter are probably both a consequence of the spiral arms (i.e. the density and velocity dispersions of the spiral arms) being better resolved. Generally though the simulations with 4 and 8 million particles are more similar than those with 1 and 4 million particles. For the runs with a less steep rotation curve (and no spiral arms) there are minimal differences at different resolutions, since there is little substructure and no reversals of the field. 

Figure~\ref{fig:bresolution} shows the evolution of the strength of the magnetic field at different resolutions, displaying $|\boldsymbol{B}|$ and $\boldsymbol{B}_{\theta}$. In all cases the evolution of the magnetic field is in much better agreement between the simulations with 4 and 8 million particles, whilst the evolution for the simulation with 1 million particles can be quite different. In addition the evolution of $B_{\theta}$ does not indicate a reversal with 1 million particles compared to the  higher resolution simulations, whereas the reversal occurs almost at the same time with 4 and 8 million particles. The evolution of $\boldsymbol{B}_{\theta}$ is also more similar with 4 and 8 million particles, with a more gradual increase, again indicative that over-amplification may be associated with numerical effects. Overall the simulations indicate some degree of convergence above 1 million particles, whilst simulations with less than 1 million particles appear likely to give erroneous results. The large difference seen in the simulation with only 1 million particles, compared to those with 4 and 8 million, is likely a consequence of the shock not being well resolved with the lower resolution.
The calculations in \citet{Stas2015}, who also modelled a galaxy using a divergence cleaning SPMHD method, used only $3.9\times10^4$ SPH particles.  

\subsection{Divergence cleaning method}
To examine the impact of the divergence cleaning on our results, we also ran a simulation with the effective speed of the cleaning four times higher (MHDN4OC4). Figure~\ref{fig:cleaning} compares this model to our fiducial model. 
The figure shows that the higher over-cleaning does not make any difference to the structure of the disc or magnetic field up to this point in the simulation, and thus the results are not dependent on over-cleaning or the magnetic divergence.  Both simulations indicate a similar reversal in the magnetic field, starting at a radius of about 3--4 kpc. 

\begin{figure}
\centerline{\includegraphics[width=0.7\columnwidth]{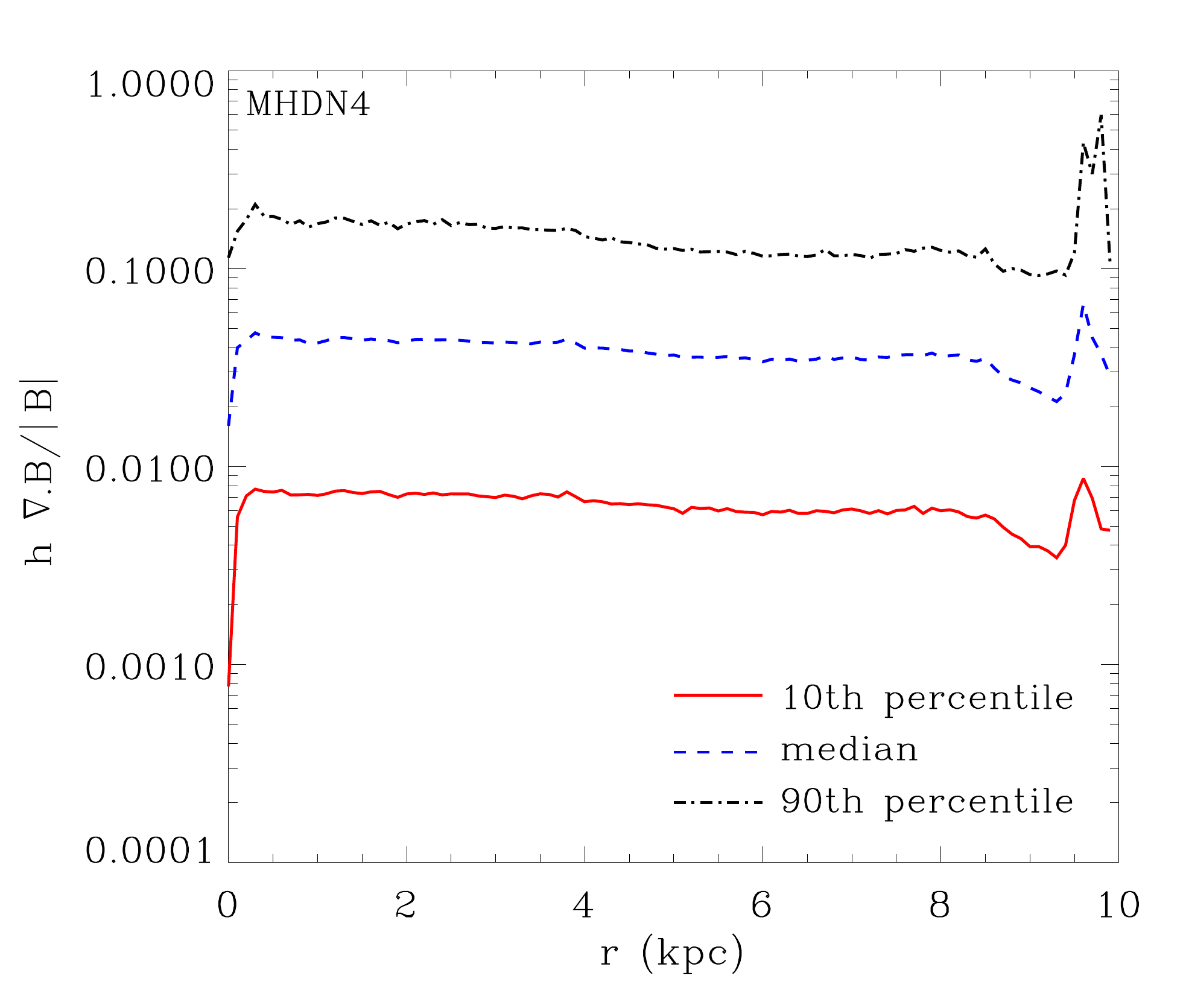}}
\centerline{\includegraphics[width=0.7\columnwidth]{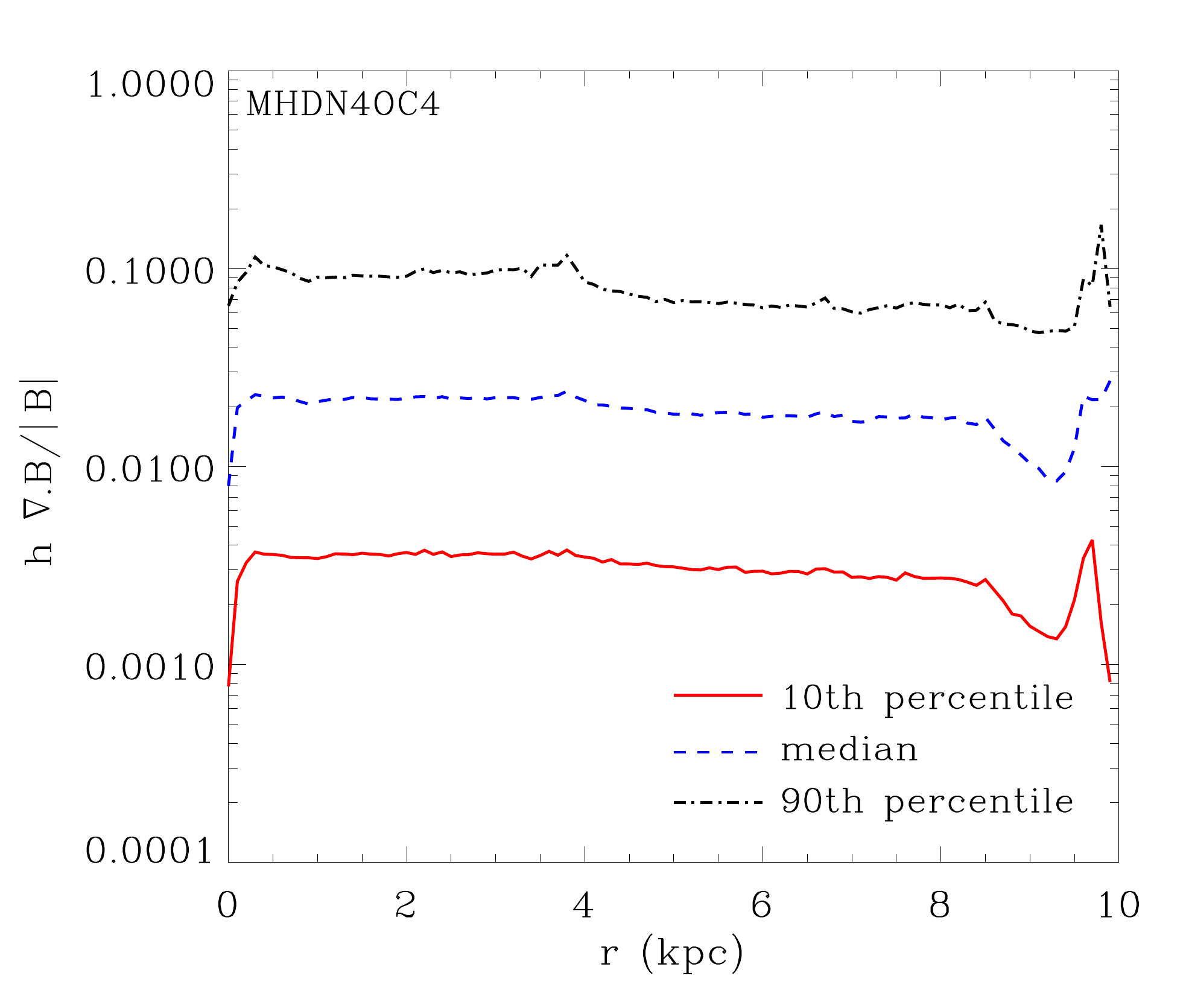}}
\centerline{\includegraphics[width=0.7\columnwidth]{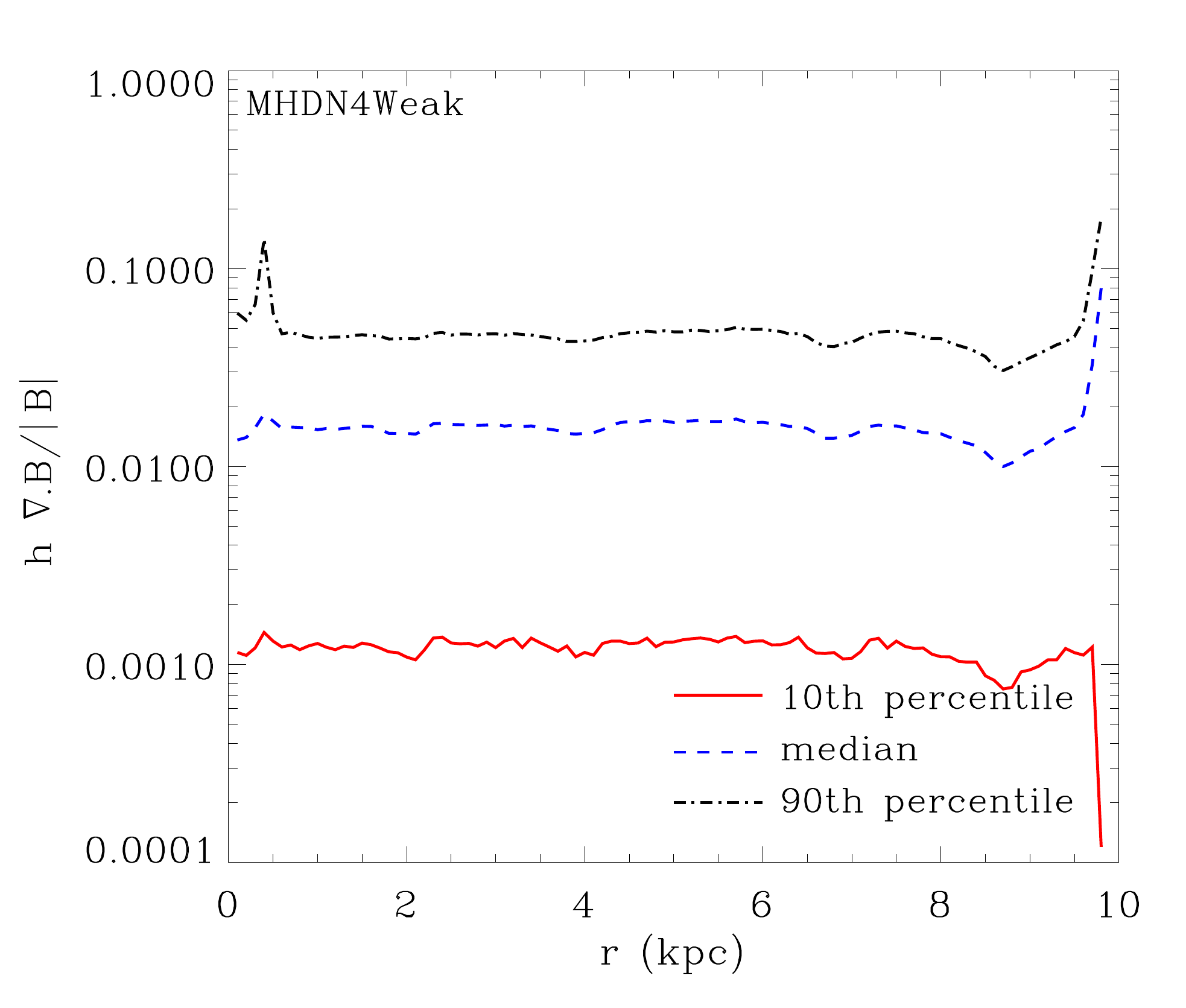}}
\caption{Divergence error, $h \vert\nabla\cdot\mathbf{B}\vert/|\mathbf{B}|$ versus radius in our fiducial simulation (top), the calculation with higher over-cleaning (centre) and the calculation with the weaker spiral potential (lower). The 10 and 90th percentiles, and median values are shown for each case at a time of 226 Myr. $h \vert\nabla\cdot\mathbf{B}\vert$/$|\mathbf{B}|$ is typically around 0.01 but is slightly lower for the simulations with higher over-cleaning and weaker spiral arms compared to the fiducial simulation.}\label{fig:hdivb}
\end{figure}

We examine how well the divergence cleaning method is limiting the divergence of the magnetic field using the dimensionless quantity $h \vert\nabla\cdot\boldsymbol{B}\vert$/$|\boldsymbol{B}|$. Figure~\ref{fig:hdivb} shows the median, 10th and 90th percentiles of this quantity on the particles binned by radius for runs MHDN4, MHDN4OC4 and MHDN4Weak at a time of 226 Myr. Ideally values of $h \vert\nabla\cdot\boldsymbol{B}\vert$/$|\boldsymbol{B}|$ should be $\lesssim 0.1$ to indicate that the divergence of $\boldsymbol{B}$ is low and that errors in the calculation of the magnetic field are minimal. Figure~\ref{fig:hdivb} indicates that $h \vert\nabla\cdot\boldsymbol{B}\vert$/$|\boldsymbol{B}|$ tends to lie between 0.001 and 0.1 to 0.01 to 0.2 depending on the calculation. In our fiducial calculation (MHDN4, top panel) $h \vert\nabla\cdot\boldsymbol{B}\vert$/$|\boldsymbol{B}|$ is highest at the edge of disc. Not long after this time, the simulation cannot be continued because the magnetic divergence becomes too high. The calculation with the higher over cleaning (middle panel) lowers the divergence compared to the fiducial calculation, both throughout the disc and the peak at the edge of the disc, although as shown in Figure~\ref{fig:cleaning} there is minimal difference in the evolution of the magnetic field at this point. This simulation can however be run further (until 260 Myr) without encountering problems with $\nabla\cdot\boldsymbol{B}$. With a weaker spiral potential (lower panel) $h \vert\nabla\cdot\boldsymbol{B}\vert /\vert\boldsymbol{B}\vert$ is lower throughout the disc, and again the simulation can be run further (another 100 Myr). The simulations with no spiral arms (MHDN4Nosp) and a shallower rotation curve (MNDN4Rc2) are similar to the weaker potential case, except the peak at the edge of the disc is not present, and again these simulations can be run further. 

We also compared a measure of the shear velocity to the Alfven speed for the models with different galactic potentials and different over-cleaning. These should give an indication of the timescales for the shear and magnetic field to evolve. The shear velocity is calculated by $A h$ where $A$ is the Oort's constant A and is a measure of shear in galaxies, and $h$ is the smoothing length. The Alfven speed is 
\begin{equation}
v_A=\frac{|\boldsymbol{B}|}{\sqrt{\mu_0 \rho}}.
\end{equation}
In all cases, the shear velocity is typically an order of magnitude lower than the Alfven speed except at the edge of the disc, indicating that typically the divergence cleaning is able to operate within the dynamical timescale imposed by the rotation curve of the galaxy. There is again a slightly higher difference between the two measures with higher over cleaning, or weak or no spiral arms.

\section{Conclusions}
We have performed SPMHD calculations of spiral galaxies with a fixed spiral potential. Our calculations adopt simple physics, assuming an isothermal ISM, and do not include self gravity, star formation or stellar feedback, but we vary the form of the galactic and spiral potential. We observe a clear dependence of the morphology of the magnetic field on the form of the spiral potential, in particular the presence of magnetic field reversals. We find that reversals of the magnetic field tend to be associated with large velocity changes across the spiral shock. We find reversals only occur if the velocity difference is $\gtrsim20$ km s$^{-1}$. The locations of the reversal in the disc are shown to be close to the location of the maximum change in velocity. This is where the magnetic field experiences the maximum distortion by the velocity field, and where the consequent straightening of the field leads to a reversal. Simulations with only weak, or no spiral arms do not show reversals, or at least not until times later than we examine. We tested our findings with the grid code {\sc Athena}, and although not completely comparable, we saw reversals occur at similar locations to with SPMHD, and with similar dependence on spiral arm strength.

The location of reversals, and large velocity changes, tend to be coincident with the ILR, as noted previously by \citet{Linden1998}. However we do not rule out reversals due to large changes in velocity not coincident with resonances, e.g. at the centre of the disc, or due a perturber or collision at larger radii. Furthermore, although we have only examined the case of a fixed spiral potential, we would expect our idea that large velocity changes induce reversals to also be true in simulations with transient spiral arms, if there is similarly a large velocity gradient. In addition to the reversals typically seen near the ILR, we also see a reversal near corotation in our model MHDN4High$\Omega$, the only model where corotation is within the simulation domain. At corotation, even though there is no spiral shock (and no change in velocity), the velocity field reverses moving from inside to outside corotation, and so a reversal is not surprising. We also made a simple comparison of the location of our reversal with the Milky Way. Both our fiducial simulation and the observations of the Milky Way show a reversal between the Sun and the centre of the Galaxy, but we caution that the dynamics of the Milky Way are not well known, and we do not know how well our simple galaxy model resembles the real Galaxy.

We also examined amplification of the field. We see significant amplification of the field, by a factor of 10 or more, particularly in the centre of the disc. Theoretical predictions indicate a linear increase in the field, whereas we see a superlinear increase in the centre. Furthermore our results with {\sc Athena} show only a small (typically $\lesssim$ 10 \%) increase in the field strength. Thus some of the amplification of the field appears to be numerical with SPMHD.

Neither field amplification nor field reversals were seen in our simulations with Euler potentials \citep{DP2008} where the field evolution is limited. Other vector potential methods may avoid the problem of over amplification of the field, whilst also allowing phenomena such as reversals, but these methods may present other problems compared to the divergence cleaning method presented here \citep{Price2010}.

Comparisons of our results with and without magnetic fields suggest that magnetic fields have only a minor effect on the disc structure, merely smoothing out substructure in the disc similar to an extra pressure term. This is a similar conclusion to our previous calculations with Euler potentials, and broadly similar to other work \citep{Lee2014,Kim2015}. However we note that our simulations are simplified, and we do not consider for example the formation or collapse of molecular clouds by self gravity. Similar also to our work with Euler potentials, the field tends to be slightly more ordered, and stronger in the spiral arms, and more random and weaker in the inter-arm regions.

We have also performed a resolution study and examined the effect of the divergence cleaning method. We find that resolution is important in these simulations, in particular we conclude that 1 million particles in a global simulation is not sufficient to obtain reliable results, but that our simulations with 4 and 8 million particles are much more consistent. We checked how well our divergence cleaning method works, and find that typically $h \vert \nabla\cdot\mathbf{B}\vert/\vert\mathbf{B}\vert$ is $\sim 0.01$ for our calculations, thus they are not affected by erroneously high values of $\nabla\cdot\mathbf{B}$. Furthermore our results are independent of the strength of the damping of the divergence. We do see edge effects start to develop at later times, but these can be diminished with stronger divergence cleaning.

\appendix
\section{Grid code tests}
We also ran tests using the {\sc Athena} grid-based code \citep{Stone2008}. We set up similar calculations of a galactic disc to those described above. We set up a cylindrical grid in {\sc Athena}  between radii of 1 and 20 kpc, -2 to 2 kpc in the vertical direction and 0 to 2$\pi$ in the azimuthal direction, using the cylindrical coordinate implementation \citep{Skinner2010}. The grid comprises 256 cells in the radial and azimuthal directions, and 64 cells in the vertical direction. We initially modelled a disc with a smaller radial extent, however we were concerned about edge effects, and a larger radial grid also allowed us to test whether a reversal occurs at corotation, which lies further out in the disc. We do not aim to carry out a direct code comparison, and particularly as our configuration
with {\sc Athena} does not allow mesh refinement, we do not achieve such high resolution with {\sc Athena} (matching the resolution between SPH and grid codes is non-trivial in any case \citep{PF2010,Few2016}). We use the same galactic, and spiral potential as run MHDN4, except the strength of the potential is higher by a factor of 3. This was in order to achieve a higher density increase in the shock, which was otherwise lower compared to {\sc sphNG} (see again \citealt{Few2016}). Indeed if we increase the resolution with {\sc Athena} we see a stronger density contrast. We adopt the same temperature of 100 K for the gas. We also ran a number of tests with different temperatures and potential strengths for comparison. As mentioned in Section~\ref{sec:amp}, we see a small amplification of the field, but here we concentrate on reversals.

\begin{figure} 
\centerline{\includegraphics[width=\columnwidth]{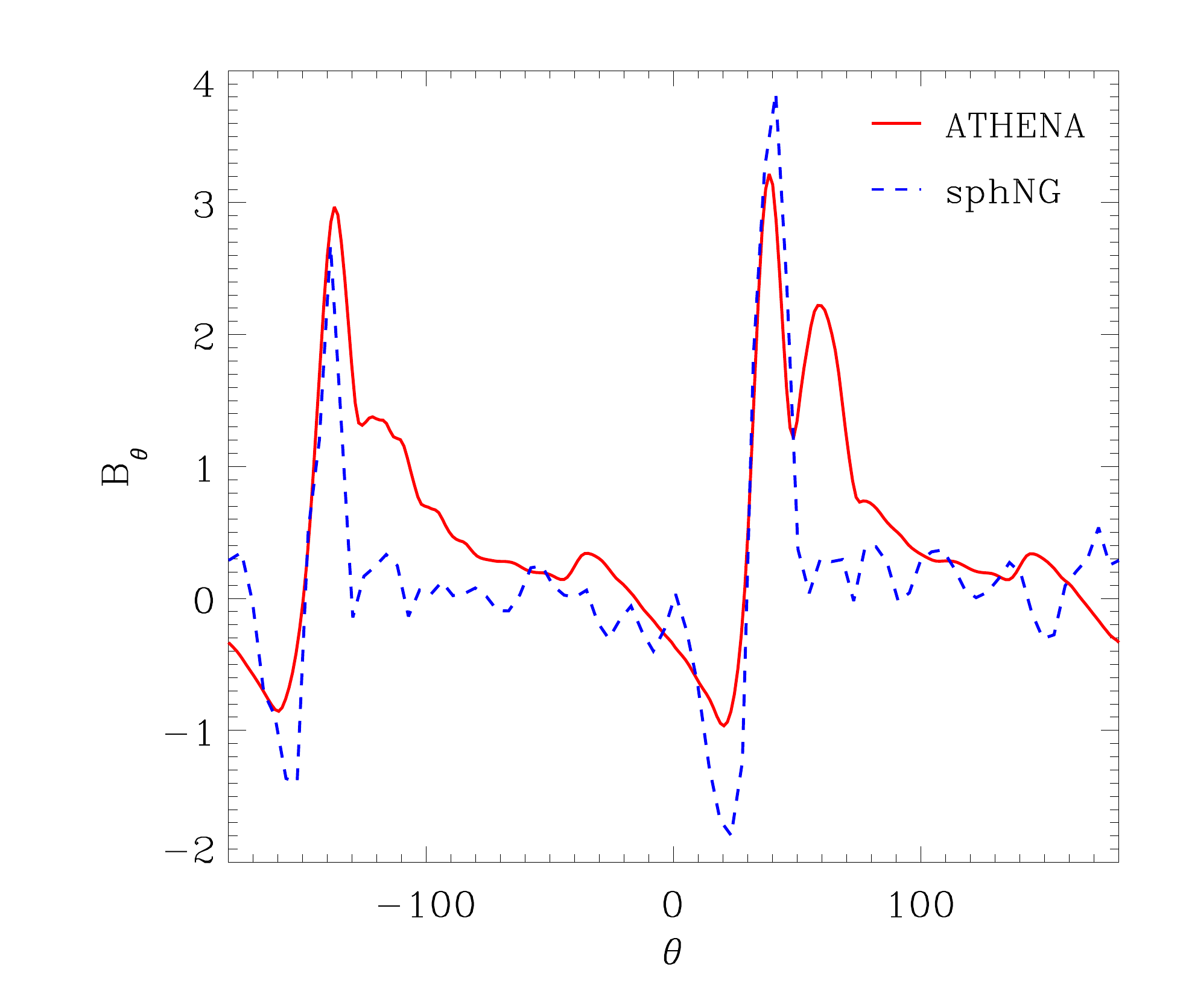}}
\caption{The azimuthal component of the magnetic field is shown at a galactic radius of 5.5 kpc for {\sc Athena} and {\sc sphNG}. The results with both codes indicate the presence of reversals. The \textsc{Athena} result is shown at a scale height of 220 pc, where the reversal is clearest (the vertical resolution is $\sim$60 pc), but the reversal is present from the midplane up to a height of 1.3 kpc. Typically the reversals for {\sc Athena} are smaller than {\sc sphNG}, and smaller than that shown in the above figure.}\label{fig:athena}
\end{figure}

From our nominal {\sc Athena}  test we see occurrences where the azimuthal component is negative, indicative of a reversal. We see reversals occurring at radii of around 2--6 kpc and 10--10.5 kpc. The first range of radii is similar to the range seen in {\sc sphNG}, which we attribute  to the ILR. The second range corresponds roughly with corotation, which again is in agreement with the {\sc sphNG} results in Section~\ref{sec:ILR}. We show a comparison of azimuthal profiles between {\sc sphNG} and {\sc Athena}  in Figure~\ref{fig:athena}. The magnetic field is shown at a radius  of 5.5 kpc versus azimuth in the top figure, at a time of 240 Myr for the {\sc Athena} simulation, and 226 Myr for {\sc sphNG}.
Since the magnetic field is higher for {\sc sphNG}, we adjust the amplitude of the magnetic field so that it is similar for both calculations (thus there are no units in the figure). In both {\sc sphNG} and {\sc Athena}, the reversal occurs at a similar location, just before the spiral arm. The reversal is weaker in {\sc Athena} compared to {\sc sphNG}, though generally the reversals in either code are not particularly strong.  
The velocity difference for the {\sc Athena} run shown in Figure~\ref{fig:athena} is $\sim$ 20 km s$^{-1}$,  so similar to the {\sc sphNG} models with reversals.

We also ran a few further tests, varying the resolution and strength of the spiral potential. We did not see large differences with resolution, but the density in the shock was higher with increased resolution. Similar to the results presented with {\sc sphNG}, reversals occur earlier with a stronger spiral potential or shock, and with a weak shock, or no spiral arms, we do not observe any reversal.

There are a few caveats to our {\sc Athena} results. One caveat is that the code often does not run much further than the results shown here, only a few Myr (though the reversals themselves start at around 200 Myr). It was not clear why this was the case, or if the reversal would be stronger at later times. The {\sc sphNG} code also had issues with following the magnetic field for longer. A second caveat is the influence of boundary conditions. We see reversals at the edge of the grid in the radial direction, hence we extend the grid significantly beyond our region of interest. 
Thirdly we do not always see reversals in the midplane of the disc for the 2--6 kpc radial range --- in fact, the location of the strongest reversal varies with vertical height at different radii (see Figure~\ref{fig:athena}).  Again it is not clear why this is, possibly large motions in the vertical direction due to spiral shocks \citep{Kimz2006} could be relevant.

Overall, the reversals are less substantial with {\sc Athena} compared to {\sc sphNG}, but they do still appear, and exhibit behaviour consistent with the {\sc sphNG} results (in terms of location and dependence on the spiral potential strength).

\section{Acknowledgments}
We thank the referee for a constructive report. We thank Jim Pringle for useful criticism.
This work used the DiRAC Complexity system, operated by the University of Leicester IT Services, which forms part of the STFC DiRAC HPC Facility (www.dirac.ac.uk ). This equipment is funded by BIS National E-Infrastructure capital grant ST/K000373/1 and  STFC DiRAC Operations grant ST/K0003259/1. DiRAC is part of the National E-Infrastructure.
This work also used the University of Exeter Supercomputer, a DiRAC Facility jointly funded by STFC, the Large Facilities Capital Fund of BIS, and the University of Exeter.
CLD acknowledges support by the European Research Council under the European Community's Seventh Framework Programme (FP7/2011-2016 grant agreement no. 280104, LOCALSTAR). DJP is funded by a Future Fellowship (FT130100034) from the Australian Research Council (ARC).
MRB and TST acknowledge support by the European Research Council under the European Community's Seventh Framework Programme (FP7/2007-2013 grant agreement no. 339248). 
\bibliographystyle{mnras}
\bibliography{Dobbs}

\bsp
\label{lastpage}
\end{document}